\input harvmac
\input amssym.tex
\overfullrule=0pt

%
\def\tilde{\widetilde}
\def\bar{\overline}

\def\pl{Phys. Lett. }

\def\cmp{Commun. Math. Phys. }

\font\zfont = cmss10 

\def\bigone{\hbox{1\kern -.23em {\rm l}}}
\def\ZZ{\hbox{\zfont Z\kern-.4emZ}}

\def\Dirac{D\hskip-.65em /}
\def\Diracs{D\hskip-.55em /}
\def\DiracB{{\cal D\hskip-.65em /}\,}
\def\flux{G\hskip-.65em /}

\def\hslash{H\hskip-.70em /}

\def\torsion{\,{\rm h}\hskip -.69em {\rm l}\,\,}
\def\ltorsion{\,{\rm h}\hskip -.54em {\rm l}\,\,}

\def\CL {{\cal L}}

\def\CO {{\cal O}}

\def\CT {{\cal T }}


\def\IZ{\relax\ifmmode\mathchoice
{\hbox{\cmss Z\kern-.4em Z}}{\hbox{\cmss Z\kern-.4em Z}}
{\lower.9pt\hbox{\cmsss Z\kern-.4em Z}}
{\lower1.2pt\hbox{\cmsss Z\kern-.4em Z}}\else{\cmss Z\kern-.4em
Z}\fi}
\def\IB{\relax{\rm I\kern-.18em B}}
\def\IC{{\relax\hbox{$\inbar\kern-.3em{\rm C}$}}}
\def\ID{\relax{\rm I\kern-.18em D}}
\def\IE{\relax{\rm I\kern-.18em E}}
\def\IF{\relax{\rm I\kern-.18em F}}
\def\IG{\relax\hbox{$\inbar\kern-.3em{\rm G}$}}
\def\IGa{\relax\hbox{${\rm I}\kern-.18em\Gamma$}}
\def\IH{\relax{\rm I\kern-.18em H}}
\def\II{\relax{\rm I\kern-.18em I}}
\def\IK{\relax{\rm I\kern-.18em K}}
\def\IP{\relax{\rm I\kern-.18em P}}

\def\IQ{\relax\hbox{$\inbar\kern-.3em{\rm Q}$}}
\def\IP{\relax{\rm I\kern-.18em P}}

\def\IB{\relax{\rm I\kern-.18em B}}
\def\IC{\Bbb{C} }
\def\ID{\relax{\rm I\kern-.18em D}}
\def\IE{\relax{\rm I\kern-.18em E}}
\def\IF{\relax{\rm I\kern-.18em F}}
\def\IG{\relax\hbox{$\inbar\kern-.3em{\rm G}$}}
\def\IGa{\relax\hbox{${\rm I}\kern-.18em\Gamma$}}
\def\IH{\relax{\rm I\kern-.18em H}}
\def\II{\relax{\rm I\kern-.18em I}}
\def\IJ{\relax{\rm I\kern-.18em J}}
\def\IK{\relax{\rm I\kern-.18em K}}
\def\IL{\relax{\rm I\kern-.18em L}}

\def\IN{\relax{\rm I\kern-.18em N}}
\def\IO{\relax{\rm I\kern-.18em O}}
\def\IP{\relax{\rm I\kern-.18em P}}
\def\IQ{\relax\hbox{$\inbar\kern-.3em{\rm Q}$}}
\def\IR{\relax{\rm I\kern-.18em R}}
\def\IW{\relax\hbox{$\inbar\kern-.3em{\rm W}$}}



\def\inbar{\,\vrule height1.5ex width.4pt depth0pt}

\font\cmss=cmss10 \font\cmsss=cmss10 at 7pt
\def\IR{\relax{\rm I\kern-.18em R}}

\def\Tr{\rm Tr}

\def\IC{{\relax\hbox{$\inbar\kern-.3em{\rm C}$}}}

\Title{ } {\vbox{\centerline{Flux corrections to anomaly
cancellation}
\smallskip
\centerline{in   M-theory on a manifold with boundary} }}
\smallskip
\centerline{Sergio Lukic and Gregory Moore}
\smallskip
\centerline{\it Department of Physics and Astronomy, Rutgers University}
\centerline{\it Piscataway, NJ 08855-0849, USA}

\bigskip
We show how the coupling of gravitinos and gauginos to fluxes
modifies anomaly cancellation in M-theory on a manifold with
boundary. Anomaly cancellation continues to hold, after a shift of
the definition of the gauge currents by a local gauge invariant
expression in the curvatures and $E_8$ fieldstrengths. We compute
the first nontrivial such correction.

Warning: Ian Moss has called into question several of
the numerical coefficients in the extended Dirac operators in
this paper. We have not confirmed this but the reader is warned
not to trust the precise coefficients in the formulae for
the Dirac operators and heat kernel expansions.  We believe these
possible errors do not change our qualitative conclusions. One of
us intends to return to the issue and recheck the formulae.
We thank Ian Moss for pointing out these problems.

\bigskip

\noindent


\Date{November 13, 2013}

\lref\sugra{E. Cremmer, B. Julia and J. Scherk, {\it Supergravity
Theory in 11 dimensions}, \pl {\bf 76B} 4 409-412 (1978).
}%
\lref\HWI{P. Ho\v rava and E. Witten, {\it Eleven-dimensional
supergravity on a manifold with boundary}, {\tt hep-th/9603142}
(1996).
}%
\lref\HWII{P. Ho\v rava and E. Witten, {\it Heterotic and type I
string dynamics from eleven-dimensions}, {\tt hep-th/9510209}
(1995).
}%
\lref\FM{D. S. Freed and G. W. Moore, {\it Setting the Quantum
Integrand of M-theory}, {\tt hep-th/0409135} (2004).
}%
\lref\DFM{E-D. Diaconescu, D. S. Freed and G. W. Moore, {\it The
M-theory $3$-form and $E_{8}$ gauge theory}, {\tt hep-th/0312069}
(2003).
}%
\lref\GSWII{M. B. Green, J. H. Schwarz and E. Witten, {\it
Superstring Theory vol.2}, Cambridge University Press (1987).
}%
\lref\bismutfreedII{J-M. Bismut and D. S. Freed, {\it The Analysis
of Elliptic Families. I. Metrics and Connections on Determinant
Bundles}, Commun. Math. Phys. {\bf 106} 159-176 (1986).
}%
\lref\bismutfreedI{J-M. Bismut and D. S. Freed, {\it The Analysis of
Elliptic Families. II. Dirac Operators, Eta Invariants, and the
Holonomy Theorem}, Commun. Math. Phys. {\bf 107} 103-163 (1986).
}%
\lref\bismut{J-M. Bismut, {\it The Atiyah-Singer index theorem for
families of Dirac operators: Two heat equation proofs},
Invent. math. {\bf 83} 91-151 (1986).
}%
\lref\finitetimeI{J. de Boer, B. Peeters, K. Skenderis and P. van
Nieuwenhuizen, {\it Loop calculations in quantum-mechanical
non-linear sigma models}, Nucl. Phys. {\bf B446} 211 (1995), {\tt
hep-th/9504097}.
}%
\lref\finitetimeII{J. de Boer, B. Peeters, K. Skenderis and P. van
Nieuwenhuizen, {\it Loop calculations in quantum-mechanical
non-linear sigma models with fermions and applications to
anomalies}, Nucl. Phys. {\bf B459} 631 (1996), {\tt hep-th/9509158}.
}%
\lref\finitetimeIII{K. Peeters and A. Waldron, {\it Spinors on
manifolds with boundary: APS index theorems with torsion}, {\tt
hep-th/9901016} (1999).
}%
\lref\FHSS{D. Fliegner, P. Haberl, M. G. Schmidt and C. Schubert,
{\it The Higher Derivative Expansion of the Effective Action by the
String Inspired Method. Part II.}, {\tt hep-th/9707189} (1997).
}%
\lref\vandeven{A. E. M. van de Ven, {\it Index-free Heat Kernel
Coefficients}, {\tt hep-th/9708152} (1997).
}%
\lref\zuk{J. A. Zuk, {\it Nucl. Phys.} {\bf 280}, p. 125 (1987).
}%
\lref\mmanual{D. V. Vassilevich, {\it Heat kernel expansion: user's
manual}, {\tt hep-th/0306138} (2003).
}%
\lref\mavromatos{N. E. Mavromatos, {\it A note on the Atiyah-Singer
index theorem for manifolds with totally antisymmetric $H$ torsion},
J. Phys. A: Math. Gen. {\bf 21} 2279-2290 (1988).
}%
\lref\obukhov{Y. N. Obukhov, {\it Spectral geometry of the
Riemann-Cartan space-times}, Nucl. Phys. {\bf B212} 237-254 (1983).
}%
\lref\witten{E. Witten, {\it On Flux quantization in M-theory and
the effective action}, {\tt hep-th/9609122} (1996).
}%
\lref\gaume{L. Alvarez-Gaum\'e, {\it Supersymmetry and the
Atiyah-Singer Index theorem}, \cmp {\bf 90} 161-173 (1983). }
\lref\gaumewitten{L. Alvarez-Gaum\'e and E. Witten, {\it
Gravitational Anomalies}, Nuclear Physics {\bf B234} 269 (1984). }
\lref\friedan{D. Friedan and P. Windey, {\it Supersymmetric
derivation of the Atiyah-Singer Index and the Chiral Anomaly},
Nuclear Physics {\bf B235} 395-416 (1984).
}%
\lref\odd{L. Alvarez-Gaume, S. Della Pietra and G. Moore, {\it
Anomalies in Odd Dimensions}, Annals of Physics {\bf 163} 288-317
(1985).
}%
\lref\bardeenzumino{W. A. Bardeen and B. Zumino, {\it Consistent and
Covariant Anomalies in Gauge and Gravitational Theories}, Fermilab
PUB-8438 T (1984).
}%
\lref\penrspin{R. Penrose, W. Rindler, {\it Spinors and Space-Time}
Vol.2., Cambridge University Press, 440-464 (1986).
}%
\lref\harmspin{N. Hitchin, {\it Harmonic Spinors}, Advances in
Mathematics, {\bf 14} 1 (1974).
}%
\lref\quillen{D. Quillen, {\it Superconnections and the Chern
Character}, Topology {\bf 24}, 89-95 (1985).
}%
\lref\getzler{E. Getzler, {\it The Bargmann representation,
generalized Dirac operators and the index of pseudodifferential
operators on $R^n$}, 63-81, Contemp. Math. {\bf 179}, AMS, (1994).
}%
\lref\moss{I. G. Moss, {\it Boundary terms for supergravity and
heterotic M-theory}, {\tt hep-th/0403106} (2004). } \lref\mossII{I.
G. Moss, {\it A new look at anomaly cancellation in heterotic
M-theory}, {\tt hep-th/0508227} (2005). } \lref\nekrasov{N.
Nekrasov, {\it Z Theory}, {\tt hep-th/0412021} (2004). }
\lref\polchinski{J. Polchinski, {\it Open Heterotic Strings}, {\tt
hep-th/0510033} (2005). } \lref\macfarlane{A. J. Macfarlane and A.
J. Mountain, {\it Construction of supercharges for the
one-dimensional supersymmetric non-linear sigma model }, {\tt
hep-th/9703024} (1997). }

\lref\scholl{M. Scholl , PhD Thesis, Univ. of Texas at Austin, 1995}

\lref\ASanomaly{M. F. Atiyah and I. M. Singer, ``Dirac operators
coupled to vector potentials,'' Proc. Nat. Acad. of Sci. USA, {\bf
81}(1984)2597. }

\lref\WittenXE{
  E. Witten,
  ``Global Gravitational Anomalies,''
  Commun.\ Math.\ Phys.\  {\bf 100}, 197 (1985).
}

\lref\FreedZX{
  D. S. Freed,
  ``Determinants, Torsion, and Strings,''
  Commun.\ Math.\ Phys.\  {\bf 107}, 483 (1986).
}

\lref\MooreJV{
  G. W. Moore,
  ``Anomalies, Gauss laws, and page charges in M-theory,''
  Comptes Rendus Physique {\bf 6}, 251 (2005)
  [arXiv:hep-th/0409158].
}

\lref\dewitt{B. S. DeWitt, {\it Quantum Field Theory In Curved
Space-Time }, Physics Reports {\bf 19}, pp. 295-357, (1975). }

\lref\BarVilko{A.O. Barvinsky and G.A. Vilkovisky, {\it The
generalized Schwinger-Dewitt technique in gauge theories and quantum
gravity}, Physics Reports {\bf 119}, (1985). }

\lref\seeley{R. T. Seeley, {\it Complex powers of an elliptic
operator}, Proc. Symp. Pure Math, (1967). }

\lref\fujikawa{K. Fujikawa, {\it Path integral for gauge theories
with fermions}, Physical Review D {\bf 21} vol.10, pp 2848-2858
(1980). }

\lref\quillen{D. Quillen, `` Determinants of Cauchy-Riemann
operators on Riemann surfaces,'' Functional Anal. Appl. 19 (1985),
no. 1, 31--34.}

\lref\AlvarezGaumeIG{
  L. Alvarez-Gaume and E. Witten,
  ``Gravitational Anomalies,''
  Nucl.\ Phys.\ B {\bf 234}, 269 (1984).
}
\lref\AlvarezGaumeDR{
  L. Alvarez-Gaume and P. H. Ginsparg,
  ``The Structure Of Gauge And Gravitational Anomalies,''
  Annals Phys.\  {\bf 161}, 423 (1985)
  [Erratum-ibid.\  {\bf 171}, 233 (1986)].
}

\lref\MooreWS{
  G. W. Moore and P. C. Nelson,
  ``The Etiology Of Sigma Model Anomalies,''
  Commun.\ Math.\ Phys.\  {\bf 100}, 83 (1985).
}

\lref\KalloshDE{
  R. E. Kallosh,
  ``Modified Feynman Rules In Supergravity,''
  Nucl.\ Phys.\ B {\bf 141}, 141 (1978).
}

\lref\NielsenMP{
  N. K. Nielsen,
  ``Ghost Counting In Supergravity,''
  Nucl.\ Phys.\ B {\bf 140}, 499 (1978).
}

\lref\MavromatosRU{
  N. E. Mavromatos,
   ``A Note On The Atiyah-Singer Index Theorem For Manifolds With Totally
  Antisymmetric H Torsion,''
  J.\ Phys.\ A {\bf 21}, 2279 (1988).
}

\newsec{Introduction and Conclusion}

M-theory on a manifold with boundary exhibits some extraordinary
features, first noted by Horava and Witten \refs{\HWI,\HWII}. First
among these features is a subtle anomaly cancellation, requiring the
presence of an independent $E_8$ super-Yang-Mills multiplet (of
either chirality) on each boundary component. In general, anomaly
cancellation is best addressed in the geometric framework of
determinant line bundles with connection
\refs{\ASanomaly,\AlvarezGaumeIG,\quillen,\MooreWS,\WittenXE,\AlvarezGaumeDR,\FreedZX}.
For recent discussions see, for example \refs{\FM,\MooreJV}. This
framework is conceptually clear, is the best approach to
cancellation of global anomalies, and is in any case the basis for
the descent formalism. In a word, it states that the effective
action after integrating out fermions must be a section of a
geometrically trivialized line bundle, that is, a topologically
trivial line bundle with a trivial connection.

Anomaly cancellation in M-theory was discussed in the geometric
framework in \FM. The present paper begins to fill a gap left open
in \FM\ and indeed left open in the entire literature on anomaly
cancellation in 10- and 11-dimensional supergravities. Namely, in
\FM\ the coupling of gravitinos and gauginos to fluxes was omitted.
In this paper it will be retained. The natural connection on a
determinant line bundle for an operator $D$ is a regularized version
of ${\Tr} D^{-1} \delta D$. Therefore,  including couplings to the
flux results in a change in the connection on the determinant line
bundle and hence in the curvature. In \FM\ it was shown that if we
omit these couplings then there is a canonical geometrical
trivialization (termed there a canonical ``setting of the
integrand'') of the line with connection $\CL_{\rm Fermi}\otimes
\CL_{CS}$. Here the fermion effective action is a section of
$\CL_{\rm Fermi}$ while  $\CL_{CS}$ accomodates the Chern-Simons
term. (See  \DFM\ for an in depth discussion of this line bundle and
its connection). Including the couplings of the fermions to the
fluxes spoils the geometrical trivialization. Nevertheless, as we
show here, the curvature of $\CL_{\rm Fermi}\otimes \CL_{CS}$ is of
the form $\CF= {\rm d}A$ where $A$ is a globally well-defined $1$-form on
the space of (gauge-equivalence classes of) bosonic fields.
Moreover, $A$ is of the form $\int_X I_{11}$ where $I_{11}$ is local
in the fields, and $X$ is the $10$-dimensional boundary. Physically
this means that although there is a change in the anomaly polynomial
$I_{12}$, it changes by $dI_{11}$ where $I_{11}$ is gauge invariant.
There is still a physical consequence of this change - the change of
connection needed to restore geometrical trivialization  corresponds
 to a change of the definition of the gauge current. We give
an explicit formula for this change, to lowest order in fluxes and
in flat space, in equation $(4.58)$ below.

This research could be continued in several possibly fruitful
directions.  The functionals that describe the flux corrections to
the curvature of the line bundle are of the same type as the ones
which were introduced by N. Nekrasov \nekrasov\ to define actions
for fluxes on manifolds of special holonomy.   There are also
analogous corrections to the gauge current in the $SO(32)$ heterotic
string. These corrections might be relevant to the open string
sector recently proposed by J. Polchinski \polchinski. Finally, it
would be interesting to carry out a similar investigation in the
formulation by I. Moss of M-theory on a boundary \moss \mossII.
His formulation   has several advantages over that of \HWI\HWII.
There are no $\delta$-functions,  and his approach is local at each
boundary. His formulation uses different boundary conditions for the
gravitinos and does not have the
$\bar{\chi}\gamma(\iota_{\nu}G^{\partial})\chi$ term for the
gauginos, which plays an important role in our analysis.

The organization of this paper is as follows: section 2 contains a
definition of the one-loop effective action in M-theory,   taking
into acount their couplings with the flux.  We derive explicitly the
contributions from the bulk and the boundary, and thus determine the
line bundle $\CL_{\rm Fermi}$, where the exponentiated effective
action is defined. In section 3, we analyze the geometry of this
line bundle. The contribution from the boundaries yields a
non-vanishing local curvature ${\cal F}_{\rm Fermi} \in
\Omega^2({\cal T})$ for $\CL_{\rm Fermi}$. Here $\CT$ is the space
of (gauge inequivalent) bosonic field configurations. After
including the contribution of $\CL_{CS}$, the total curvature is a
globally exact form ${\cal F}={\rm d}A$. Thus, it is possible to
obtain a geometrical trivialization by changing the connection.
Similarly, the contribution from the bulk gives rise to possible
$\IZ_2$-holonomies for loops in $\pi_1({\cal T})$, due to an
ambiguity in the definition of the sign of the Rarita-Schwinger
determinant \witten\FM. We show how the flux corrections do not
alter the usual $\IZ_2$ (or parity) anomaly cancellation mechanism.
Section 4 provides explicit formulas for the curvature of the line
bundle when the boundaries of $Y$ are  flat Euclidean space. We show
how our calculations, based on heat kernel expansions and the
descent formalism, confirm the general arguments given in section 3.
  For completeness, we also study this local anomaly
using Fujikawa's method, determining the flux correction to the
gauge current as a gauge invariant 9-form in $\Omega^{9}(\IR^{10})$.
Appendix A states our Clifford algebra conventions. Appendix B
briefly indicates the connection to supersymmetric quantum
mechanics.

\newsec{The one-loop effective action}

In this section we sketch the gravitino partition function in the
case of M-theory on a spin $11$-dimensional manifold $Y$, which
might have a nonempty boundary.

 The supergravity multiplet consists of the
metric $g$, a gravitino $\psi$, and a $3$-form gauge potential with
corresponding field strength $G$. The low energy limit of M-theory
is described by $11$-dimensional supergravity \sugra. Here we focus
on  the quadratic part of the action for the gravitino
\eqn\gravi{{-1\over 2}\int_{Y}{\rm vol}(g) \Big[
\bar{\psi}_{I}\gamma^{IJK}D_{J}\psi_{K} + {\ell^3 \over 96} \big(
\bar{\psi}_{I}\gamma^{IJKLMN} \psi_{N} +
12\bar{\psi}^{J}\gamma^{KL}\psi^{M} \big) G_{JKLM}\Big]} with
$I,J,\dots$ worldindices,  $D_{I}$ the spin connection and $\ell$
the eleven dimensional Planck length. We are neglecting higher order
terms in $\psi_I$.  The local supersymmetry transformation for the
gravitino up to $3$-fermi terms, is
\eqn\susyvar{\delta\psi_{I}=D_{I}\epsilon + {\ell^3 \over
288}(\gamma_{I}\, ^{JKLM}-
8\delta^{J}_{I}\gamma^{KLM})G_{JKLM}\epsilon :=\hat{D}_{I}\epsilon.
}
We will write  \susyvar\  as $\delta\psi_{I}=\hat{D}_{I}\epsilon$,
and will refer to  $\hat{D}_{I}$   as the supercovariant derivative.
We will abbreviate the action as
\eqn\fakeact{\int_{Y}\bar{\psi}R
\psi.}
 Denote by ${\bf S}$ the spin bundle on $Y$. The generalized
Rarita-Schwinger operator $R:\Gamma(S\otimes T^{\ast}Y)\to
\Gamma(S\otimes T^{\ast}Y)$, fits into the complex \eqn\comp{0\to
\Omega^{0}({\bf S}){\buildrel \hat{D} \over \longrightarrow}
\Omega^{1}({\bf S}) {\buildrel R \over \longrightarrow}
\Omega^{1}({\bf S}) {\buildrel \hat{D}^{\ast} \over \longrightarrow}
\Omega^{0}({\bf S})\to 0, } if we require the vanishing of   $R\circ
\hat{D}$. Furthermore,   at the level of principal symbols the
complex is exact so \comp\ defines an elliptic complex. To check the
exactness of \comp\ at the level of symbols it is enough to work in
flat space, thus if $\sigma_{\hat{D}}(k)=k\in T^{\ast}Y$ is the
principal symbol associated to $\hat{D}$ and the symbol for $R$ is
$\sigma_{R}(k)=\gamma^{MNP}k_{N}$, then ${\rm Ker}(\sigma_{R}(k))$
consists of the elements ${\bf s}\sigma_{\hat{D}}(k)$ for a spinor
${\bf s}$.

The consistency condition $R\circ \hat{D}=0$ requires that the
equations of motion for the bosonic fields must be satisfied as we
show below. Hence, if we write the equations of motion for the
gravitino field as \sugra\foot{Note that we are expressing the
Rarita-Schwinger operator $R$ in two equivalent ways.}
\eqn\eqmov{R\psi=\gamma^{MNP}\hat{D}_{N}\psi_{P}=0,}
we can write
the condition $R\circ \hat{D}=0$ as
\eqn\elli{R\circ
\hat{D}=\gamma^{MNP}[\hat{D}_{N},\hat{D}_{P}]=0.}

We can describe the  bosonic configurations satisfying \elli\ by
considering the seemingly simpler relation
\eqn\ellib{\gamma^{P}[\hat{D}_{N},\hat{D}_{P}]=0.}
We claim that \elli\ and \ellib\ are equivalent.  That \ellib\
implies \eqmov\ follows from
$\gamma^{MNP}=\gamma^{M}\gamma^{N}\gamma^{P}+g^{NP}\gamma^{M}-g^{MP}\gamma^{N}+g^{MN}\gamma^{P}$.
 To prove the converse observe that
 \eqn\we{
0 =( g_{QM} + {1\over n-2}\gamma_Q \gamma_M ) \times
(\gamma^{MNP}[D_N, D_P])   = 2 \gamma^{P}[D_N, D_P]. }

By a straightforward computation  can express the condition \elli\
on the bosonic fields, using the relation \ellib\ as follows
\eqn\cala{\eqalign{\gamma^{N}[\hat{D}_{M},\hat{D}_{N}]=-{\ell^{3}\over
288} (D_{[N}G_{PQRS]})\gamma^{MNPQRS} +{5\ell^{3}\over
144}(D_{[M}G_{NPQR]})\gamma^{NPQR}\cr -{\ell^{3}\over
72}\Big(D^{N}G_{NPQR}+{\ell^{3}\over 4\times 288}G^{I_{1} \ldots
I_{4}}G^{J_{1}\ldots J_{4}} \varepsilon_{I_{1}\ldots
I_{4}J_{1}\ldots J_{4}PQR} \Big)g_{MT}\gamma^{TPQR}\cr
+{\ell^{3}\over 12}\Big(D^{N}G_{NMPQ}+{\ell^{3}\over 4\times
288}G^{I_{1}\ldots I_{4}} G^{J_{1}\ldots J_{4}}
\varepsilon_{I_{1}\ldots I_{4}J_{1}\ldots J_{4}MPQ}
\Big)\gamma^{PQ}\cr -{1\over 2}\Big( {\cal R}_{MN} - {\ell^{6}\over
6} \Big( G_{MPQR}G_{N}^{\,\, PQR} -{1\over 12}g_{MN}G_{PQRS}G^{PQRS}
\Big)\Big)\gamma^{N}=0. }} Here, we expand \ellib\ in terms of
completely antisymmetrized products of gamma matrices (see Appendix
A), hence \cala\ implies the following constraints for the bosonic
fields \eqn\eqma{dG=0} \eqn\eqmb{d\star G= -{\ell^{3}\over 2}
G\wedge G} \eqn\eqmc{{\cal R}_{MN}={\ell^{6}\over
6}\Big(G_{MPQR}G_{N}^{\,\, PQR}-{1\over 12} g_{MN}\star
(G\wedge\star G) \Big)} where ${\cal R}_{MN}$ is the Ricci tensor.
These are just the classical equations of motion of 11-dimensional
supergravity.

\subsec{The gravitino partition function.}

Since the local fermionic gauge symmetries of $n=11$ supergravity do
not close into a super Lie algebra for off-shell bosonic
backgrounds, we should in principle use the BV quantization
procedure to get a correct gauge fixed action. In this paper we
determine the gauge fixed action for backgrounds that satisfy \eqmb,
and \eqmc. This allows us to use standard BRST procedures
\KalloshDE\NielsenMP\ and simplifies the discussion considerably. Of
course, it leaves an important gap in our treatment. Accordingly, we
consider the gravitino partition function \eqn\pint{{\cal
Z}=\int_{\Omega^{1}({\bf S})/{\rm Im}\hat{D}}
[d\psi]e^{-\int_{Y}\bar{\psi}R\psi},}

It is useful to introduce the notation:
 \eqn\ea{\flux =
\gamma^{PQRN}G_{PQRN}} \eqn\eb{\flux_{N} =
\gamma^{PQR}G_{PQRN}=-\gamma^{PQR}G_{NPQR}} \eqn\eb{\flux_{RN} =
\gamma^{PQ}G_{PQRN}.}
 A direct calculation shows that
\eqn\ideqa{\gamma_{M}\, ^{PQRN}G_{PQRN}=\gamma_{M}\flux -
4\flux_{M},}
and therefore we can write \susyvar\ as
\eqn\supcova{\hat{D}_{M}\epsilon=D_{M}\epsilon+{\ell^3 \over
288}\gamma_{M}\flux\, \epsilon + {\ell^3 \over
72}\flux_{M}\epsilon.}
Since   $\gamma^{M}\gamma_{M}=-11$ and $\gamma^{M}\flux_{M}=-\flux$,
the associated supercovariant Dirac operator will be
\eqn\superdirac{\widehat{\Dirac}=\gamma^{M}\hat{D}_{M}=\Dirac-{5\ell^{3}\over
96}\flux.} Thus we can write the action \gravi\ as
\eqn\gravibb{{-1\over 2}\int_{Y}{\rm vol}(g)\Big[
\bar{\psi}_{I}\gamma^{IJK}D_{J}\psi_{K} +  {\ell^3 \over 96}
\bar{\psi}_{I}(\gamma^{IK}\flux -
8\gamma^{[I}\flux^{K]}-24\flux^{IK} )\psi_{K}\Big].}

We now use  the formal $BRST$ procedure to determine the gravitino
gauge fixed action, and choose the gauge $s=\gamma\cdot\psi$ for an
arbitrary spinor $s\in\Omega^{0}({\bf S})$. This leaves unfixed
zeromodes of the Dirac equation, constituting a finite dimensional
space  which we will deal with presently.
  Following standard procedure we write \eqn\onea{{\bf 1}=
\int_{\Omega^{0}({\bf S})^{\bot} }[{\rm d}\epsilon ]\,
\delta(s-\gamma^{M}(\psi_{M}+ \hat{D}_{M}\epsilon))({\rm det}'
\widehat{\Dirac})^{-1} } with $\Omega^{0}({\bf S})^{\bot} = \big(
{\rm Ker}\, \widehat{\Dirac} \big)^{\bot}$ and where $\hat{D}_{M}$
and $\widehat{\Dirac}$ were defined in \supcova\ and \superdirac. We
now insert \onea\ into \eqn\partition{\int [{\rm
d}\psi]e^{-\int_{Y}\bar{\psi}R\psi}} and divide by the volume of the
supergauge group to obtain the gauge-fixed expression
\eqn\partitionb{\int [d\psi]\delta(s-\gamma\cdot\psi)({\rm det}'
\widehat{\Dirac})^{-1} e^{-\int_{Y}\bar{\psi}R\psi}.} Ghost fields
are introduced by writing the determinant \onea\ in terms of
commuting ghost $\epsilon$ and antighost $\beta$ fields as
\eqn\determin{({\rm det}' \widehat{\Dirac})^{-1}=\int  [{\rm d}
\beta] [{\rm d} \epsilon] e^{-\int \bar{\beta}\widehat{\Diracs}\,
\epsilon }, } the prime in the determinant denotes the omission of
the null eigenvalues.

Furthermore we invoke the following algebraic identity for
$\phi_{M}=\psi_{M}+{1\over 2}\gamma_{M}(\gamma\cdot\psi)$, which
allows us to split the gauge fixed action as a sum of functionally
independent quadratic terms, i.e. we have the relation
\eqn\identi{-\bar{\phi}\widehat{\Dirac}_{T^{\ast}Y}\phi=
\bar{\psi}R\psi-{1\over
4}(n-2)\bar{(\gamma\cdot\psi)}\widetilde{\Dirac}\,
(\gamma\cdot\psi),} where $R$ was defined in \gravibb\ to be
\eqn\defr{R^{IK}=\gamma^{IJK}D_{J}+{\ell^3\over 96}
(\gamma^{IK}\flux - 8\gamma^{[I}\flux^{K]}-24\flux^{IK}), } while
$\widetilde{\Dirac}$ and $\widehat{\Dirac}_{T^{\ast}Y}$ are uniquely
fixed to be the generalized Dirac operators
\eqn\diracb{\widetilde{\Dirac}=\Dirac +{\ell^{3}\over 288}\flux} and
\eqn\cota{\widehat{\Dirac}_{T^{\ast}Y}=\Dirac_{T^{\ast}Y}-{\ell^{3}\over
96}\flux.}
Here, the subscript $T^{\ast}Y$ denotes the coupling with the
cotangent bundle of $Y$. The identity \identi\ is easy to check when
we substitute the $\phi$-field and the operators
$\widehat{\Dirac}_{T^{\ast}Y}$ and $\widetilde{\Dirac}$ in it and
use the following relations for $\flux$ and the gamma matrices
\eqn\wa{\eqalign{\gamma^{M}\flux - \flux\gamma^{M}=8\flux^{M}\cr
\gamma^{M}\flux^{P}+\flux^{P}\gamma^{M}=-6\flux^{MP}\cr
\gamma^{IK}=\gamma^{I}\gamma^{K}+g^{IK}.}} At this point, rather
than setting $s=0$ we average over $s=(\gamma\cdot\psi)$ using the
expression
\eqn\onea{{\bf 1}={1\over ({\rm det'}\widetilde{\Dirac})^{1/2}}
\int_{(\Omega^{0}({\bf S}))^\perp}[{\rm
ds}]\,e^{-\int\bar{s}\widetilde{\Diracs}\, s}.}
%
%

Formally, using \identi\
the gauge fixed partition function for the gravitino can be written as
\eqn\parti{{\cal Z}' = {1 \over({\rm det} \widetilde{\Dirac})^{1/2}}\int
[{\rm d}\psi][{\rm d}\beta][{\rm d} \epsilon]
{\rm exp}\Bigg( -2\pi\int_{Y}{\rm vol}(g)(
\bar{\psi}\widehat{\Dirac}_{T^{\ast}Y}\psi
-\bar{\beta}\widehat{\Dirac}\, \epsilon
)\Bigg).
}

We still must fix the remaining global fermionic symmetries given by
supercovariantly constant spinors. We will assume the procedure
described in \FM, eq. $(A.11)$ continues to hold. The net result is
the following key statement. \foot{We are unaware of an adequate
treatment of the ghost zeromodes in the gravitino partition function
in the literature. In this paper we sidestep that issue and assume
that the gravitino effective action is a section of eq. $(2.35)$.}

Let  ${\cal T}$ denote the space of bosonic  M-theory data on $Y$,
i.e., the   Riemannian metrics and $G$-fluxes, and introduce  the
fibration ${\cal Y}\to {\cal T}$  whose fiber is the spacetime
manifold $Y$. This  yields a family of operators
$(\widehat{\Dirac},\, \widetilde{\Dirac},\,
\widehat{\Dirac}_{T^{\ast}Y})$ built up fiberwise in ${\cal Y}$
through the geometric data parametrized by ${\cal T}$:
\eqn\superdirac{\widehat{\Dirac}=\Dirac-{5\ell^{3}\over 96}\flux,}
\eqn\diracb{\widetilde{\Dirac}=\Dirac +{\ell^{3}\over 288}\flux,}
\eqn\cota{\widehat{\Dirac}_{T^{\ast}Y}=\Dirac_{T^{\ast}Y}-{\ell^{3}\over
96}\flux.}
Then,   the gravitino partition function ${\rm exp}(-\Gamma_{\rm
gravitino})$ is a section of the line bundle
\eqn\newpart{ \CL_{\rm gravitino}:={\rm
Pfaff}\,\widehat{\Dirac}_{T^{\ast}Y}\otimes ({\rm Pfaff}
\widetilde{\Dirac})^{-1} \otimes ({\rm Det}\,\widehat{\Dirac})^{-1}
\to {\cal T},}
In fact, this is a line bundle with connection, as we will discuss
below. In addition, the Chern-Simons term of M-theory is also a
section of a line bundle with connection ${\cal L}_{CS}\to {\cal
T}$, and hence the M-theory measure is a section of
\eqn\total{ \CL_{\rm gravitino} \otimes \CL_{CS} \to \CT}

\subsec{Boundary contribution to the effective action.}

Let us now turn to the case where $Y$ has a boundary.
  We denote by $\partial Y_{i}$ the different
connected components and by $\omega^{\partial}$   Clifford
multiplication by the volume $10$-form on the boundary. We follow
closely the discussion of boundary conditions in \FM.  We fix a
spatial boundary condition for the spinor field $\Psi$, by imposing
\eqn\boundcond{ \epsilon_{i}\Psi^{\partial}=\Psi^{\partial}\quad
{\rm with}\,\, \epsilon_{i}=i\omega^{\partial}\,\, {\rm
or}\,\, \epsilon_{i}=-i\omega^{\partial} }
at each connected component $\partial Y_{i}$. The presence of
boundaries produces local anomalies in the theory.

The fermionic content at the boundary in the low energy description
of M-theory comes from the restriction of the gravitino and the
presence of gauginos.  We generalize the discussion of Horava and
Witten and  attach an independent $N=1$ super Yang-Mills multiplet
with gauge group $E_{8}$ and chirality $\epsilon_i$ to each
connected component of the boundary.  According to \HWI, we should
write the quadratic part of the action for the gauginos, as
\eqn\actgaug{S_{i}=-{1\over 4\pi \ell^{6}}\int_{\partial Y_{i}} {\rm
vol}(g^{\partial}) {\rm tr}\Bigg[ \bar{\chi}\Dirac_{E_{8}}\chi
-{\ell^{3}\over 24}\bar{\chi}^{a}
\gamma(\iota_{\nu}G^{\partial})\chi^{a} \Bigg]. }
The superscript $ ^{\partial}$ denotes restriction of the field on
the boundary and $\iota_{\nu}$ is the contraction with the unit
outward normal vector field to  $\partial Y_{i}$. As shorthand, we
can write the action using the generalized Dirac operator
\eqn\suprgaug{\hat{\Dirac}_{E_{8}}=\Dirac_{E_{8}}- {\ell^{3}\over
24}\gamma(\iota_{\nu}G^{\partial}),} so the exponentiated effective
action for $\chi$ is section of the line bundle
\eqn\gauginoline{
\CL_{\rm gaugino}=\bigotimes_{\epsilon_{i}=-i\omega^{\partial}}
\Big( {\rm Pfaff} \hat{\Dirac}_{E_{8}}^{\partial Y_{i}} \Big)
\bigotimes_{\epsilon_{i}=i\omega^{\partial}} \Big( {\rm
Pfaff} \hat{\Dirac}_{E_{8}}^{\partial Y_{i}}
\Big)^{-1}\longrightarrow {\cal T}, }
It is useful to decompose the boundary $4$-form $G$, in its
tangential and normal components \eqn\Gdecomp{ G^{\partial}=
\nu^{\flat}\wedge \iota_{\nu}G^{\partial}+ (1-\nu^{\flat}\wedge
\iota_{\nu})G^{\partial}= G_{N}^{\partial} + G_{T}^{\partial} } with
$\nu^{\flat}$ the $1$-form dual to the unit normal vector field
$\nu$. Also, we introduce the local ``torsion''
\eqn\defitorsion{
\torsion = -{1\over 24}\ell^{3}\iota_{\nu}G^{\partial}.}
%
%

For the gravitino sector, we have to generalize the gauge fixed
action ${\rm exp}(-\Gamma_{\rm gravitino})$ to the case  $\partial
Y\not=\emptyset$. To do this recall the relation between a
Dirac-like operator on the boundary $\partial Y_{i}$ and that  in
the bulk close to the boundary $\partial Y_{i}\times [0,\,\epsilon)$
\eqn\diracbound{ \widehat{\Dirac} =
\gamma^{\nu}(\partial_{\nu}-\widehat{\Dirac}^{\partial}) } where
$\nu$ is the normal unit vector field to the boundary. Then, as
$(\gamma^{\nu})^2=-1$, the generalized Dirac operators that we have
to study on the boundary are \eqn\gendir{\eqalign{
\widehat{\Dirac}_{T^{\ast}Y}= \Dirac_{T^{\ast}Y} - {\ell^{3}\over
96}\flux,\quad \Rightarrow\quad
\widehat{\Dirac}_{T^{\ast}Y}^{\partial} =
\Dirac_{T^{\ast}Y}^{\partial} - {\ell^{3}\over
24}\gamma(\iota_{\nu}G_{N} + \nu^{\flat}\wedge G_{T}) \cr
\widetilde{\Dirac}=\Dirac +{\ell^{3}\over 288}\flux,\quad
\Rightarrow\quad \widetilde{\Dirac}^{\partial}=\Dirac^{\partial}
+{\ell^{3}\over 72} \gamma(\iota_{\nu}G_{N} + \nu^{\flat}\wedge
G_{T}) \cr \widehat{\Dirac}= \Dirac - {5\ell^{3}\over 96}\flux,\quad
\Rightarrow\quad \widehat{\Dirac}^{\partial}= \Dirac^{\partial} -
{5\ell^{3}\over 24} \gamma(\iota_{\nu}G_{N} + \nu^{\flat}\wedge
G_{T}) }}
as $\iota_{\nu}G_{N}$ is a $3$-form and $\nu^{\flat}\wedge G_{T}$ a
$5$-form, the operators $\widehat{\Dirac}_{T^{\ast}Y}^{\partial}$,
$\widetilde{\Dirac}^{\partial}$ and $\widehat{\Dirac}^{\partial}$
anticommute with $\omega^\partial$ and hence have a well defined
index.


The restriction $\psi^{\partial}$ of the Rarita-Schwinger field
$\psi\in \Omega^{1}({\bf S})$ to $\partial Y_{i}$ decomposes into
tangential and normal components: \eqn\decomprarita{
\psi^{\partial}= \psi^{\partial}_{T}+\psi^{\partial}_{\nu} } and
their boundary conditions are given by the following definite choice
of sign \eqn\boundcondrs{\eqalign{
\omega^{\partial}\psi^{\partial}_{T}=+i\psi^{\partial}_{T}\cr
\omega^{\partial}\psi^{\partial}_{\nu}=-i\psi^{\partial}_{\nu}.
}} These boundary conditions imply that the gauge group must be
restricted by \eqn\boundcondg{\eqalign{
\omega^{\partial}\hat{\nabla}_{T}\epsilon^{\partial}=
+i\hat{\nabla}_{T}\epsilon^{\partial}\cr
\omega^{\partial}\hat{\nabla}_{\nu}\epsilon^{\partial}=
-i\hat{\nabla}_{\nu}\epsilon^{\partial}\cr }} where
\eqn\firstsusy{\eqalign{ \hat{\nabla}_{\nu} =\nu^{M}\hat{D}_{M} \cr
\hat{\nabla}_{T} = \hat{D}_{M} - \nu_{M}\hat{\nabla}_{\nu} }} and
$\hat{D}_{M}$ is the supersymmetric variation of the gravitino
\eqn\supcovabo{\hat{D}_{M}=D_{M}+{\ell^3 \over 288}\gamma_{M}\flux +
{\ell^3 \over 72}\flux_{M}.} We then choose boundary conditions on
the other ghost $\beta$, so that $\widehat{\Dirac}^{\partial}$ is
skew-adjoint: \eqn\boundcondga{\eqalign{
\omega^{\partial}\hat{\nabla}_{T}\beta^{\partial}=
-i\hat{\nabla}_{T}\beta^{\partial}\cr
\omega^{\partial}\hat{\nabla}_{\nu}\beta^{\partial}=
+i\hat{\nabla}_{\nu}\beta^{\partial}\cr }} The third ghost
that comes from integrating over $s$ \onea, has the same boundary
conditions as $\gamma\cdot\psi$: \eqn\boundconds{
\omega^{\partial}(s^{\partial})=-is^{\partial}. } As the
chiralities of $\epsilon^{\partial}$ and $\beta^{\partial}$ are
opposite, \boundcondg, and \boundcondga, lead to pfaffian line
bundles which cancel, therefore $\widehat{\Dirac}^{\partial}$ does
not appear in our analysis. On the other hand, as $\psi_{\nu}$ comes
from a component of the Rarita-Schwinger field in $\Omega^{1}({\bf
S})$, it couples to \eqn\dilatinoperator{
\widehat{\Dirac}_{\nu}^{\partial} = \Dirac^{\partial} -
{\ell^{3}\over 96}\gamma(\iota_{\nu}G_{N} + \nu^{\flat}\wedge G_{T})
}
and $s$ couples to $\widetilde{\Dirac}^{\partial}$, as  defined in
\gendir. Therefore, according to the theorems stated in \FM, based
on theorems  proved by M. Scholl \scholl,  the boundary contribution
to the exponentiated effective action ${\rm exp}(-\Gamma_{\rm
gravitino})$ is section of
\eqn\linegravitino{\eqalign{ {\cal L}_{\rm
gravitino}=\bigotimes_{-i\omega^{\partial}}\Big[ \big( {\rm
Pfaff} \widehat{\Dirac}_{T^{\ast}Y}^{\partial Y_{i}}
\big)^{1/2}\otimes\big( {\rm Pfaff} \widehat{\Dirac}_{\nu}^{\partial
Y_{i}} \big)^{-1/2} \otimes \big( {\rm Pfaff}
\widetilde{\Dirac}^{\partial Y_{i}} \big)^{-1/2} \Big] \cr
\bigotimes_{+i\omega^{\partial}}\Big[ \big({\rm Pfaff}
\widehat{\Dirac}_{T^{\ast}Y}^{\partial Y_{i}}
\big)^{-1/2}\otimes\big( {\rm Pfaff}
\widehat{\Dirac}_{\nu}^{\partial Y_{i}} \big)^{+1/2} \otimes \big(
{\rm Pfaff} \widetilde{\Dirac}^{\partial Y_{i}} \big)^{+1/2}\Big]
\to {\cal T} }}
where we are taking into account the contribution from every
connected component of the boundary. Finally we
have
\eqn\fermibundle{ \CL_{\rm Fermi} =  \CL_{\rm gaugino}\otimes \CL_{\rm
gravitino}. }

In the following sections, we  study the curvatures of the
determinant line bundles associated to generalized Dirac operators.
The $G$-dependent contributions to the curvature of \fermibundle,
are given by terms constructed with the exterior derivatives ${\rm
d}(\nu^{\flat}\wedge G_{T})$ and ${\rm d}\iota_{\nu}G_{N}$. Now
\eqn\vanish{ {\rm d}(\nu^{\flat}\wedge G_{T})= {\rm
d}\nu^{\flat}\wedge G_{T} - \nu^{\flat}\wedge {\rm d}G_{T}=0. }
To see this,  we work in the neighborhood of the boundary $\partial
Y_{i}\times [0,\, \epsilon)$  such that ${\rm d}\nu^{\flat}=0$.
Also, as $G_{T}$ is closed on the boundary we have ${\rm d}
G_{T}=0$. Thus we can neglect the contributions from
$\nu^{\flat}\wedge G_{T}$, and just work with the local torsion
$\torsion$ of \defitorsion.

\subsec{Ho\v rava-Witten reduction}

It is useful to connect our formalism to the standard   Ho\v
rava-Witten setup $Y=X\times [0,\, 1]$, used to describe the
strongly coupled heterotic string with gauge group $E_{8}\times
E_{8}$, in its low energy limit.

The H flux of heterotic string theory is recovered from the M-theory
data according to
 \eqn\deflux{{\rm H} = \int_{[0,\, 1]} {\rm d}t
G_{11\, MNP}
{\rm d}x^{M}\wedge {\rm d}x^{N}\wedge {\rm d}x^{P}
} with $t=x^{11}$ and $1\leq M,\, N,\, P\leq 10 $. On the other
hand, using the decomposition of the G-flux in terms of tangential
and normal components to the $11^{\rm th}$-coordinate
\eqn\decm{ G=
G_{11\, MNP}{\rm d}t\wedge{\rm d}x^{M}\wedge {\rm d}x^{N}\wedge {\rm
d}x^{P} + G_{QRST}{\rm d}x^{Q}\wedge{\rm d}x^{R}\wedge {\rm
d}x^{S}\wedge {\rm d}x^{T} = G_{N}+G_{T} }
 with the indices $M,N,\dots$ running
between $1$ and $10$. On the boundaries $\iota^*(G_T)$ at $t=0,1$ we
have $\iota^*_tG_{T}={\rm tr}\, F^{2}_{t} - {1\over 2}{\rm tr}\,
{\cal R}^{2}_{t}\in \Omega^{4}(X)$ where $F_t$, $t=0,1$ is the
curvature of the $E_8$ bundle on the boundary $X_t$. If we extend
$G_T$ as a family of closed forms on $X$ then
 \eqn\deriv{ 0={\rm
d}_{11}G= \Big( {\rm d}t\wedge {\partial \over
\partial t} +{\rm d}\Big) (G_{N} + G_{T})={\rm d}G_{N} + {\rm
d}t\wedge {\partial \over
\partial t} G_{T} }
($d$ and $d_{11}$ are exterior derivatives on $X$ and $Y$,
respectively). Therefore, from \deriv \eqn\equalimpor{ {\rm d}G_{N}
= -{\rm d}t\wedge {\partial \over \partial t} G_{T} }
Using \deflux\ and  \equalimpor\  we recover the usual formula
 \eqn\dhflux{ {\rm d}\,{\rm H} = {\rm tr}\,
F^{2}_{1} + {\rm tr}\, F^{2}_{2} - {\rm tr}\, {\cal R}^{2} }

Finally we would like to see how the interaction term
\eqn\actgaugg{\Delta S={1\over 96\pi \ell^{3}}\int_{X} {\rm
vol}(g_{X}) {\rm Tr}_{\bf 496}\Big[\bar{\chi} \gamma({\rm H})\chi
\Big], } in heterotic string theory, is recovered from the boundary
interactions of M-theory \actgaug \eqn\actgauggg{\Delta
S_{i}={1\over 96\pi \ell^{3}}\int_{\partial Y_{i}} {\rm
vol}(g^{\partial}) {\rm Tr}_{\bf 248}\Big[ \bar{\chi}_{i}
\gamma(\torsion)\chi_{i} \Big]. } with $i=1,\, 2$ labeling the
boundaries of the cylinder $X\times [0,\, 1]$. In the zeromode limit
we have
\eqn\constraintonG{ {\cal L}_{t}G_{N}=0,\quad {\rm or}\quad
\iota_{t}G_{N}=
\iota_{t}G_{N}^{\partial_{1}}=\iota_{t}G_{N}^{\partial_{2}}={\rm
H}=\torsion_{1}=\torsion_{2} }
i.e. $G_{N}$ is $t$-independent and the non-trivial $t$-dependence
of $G$ comes from $G_{T}$. Therefore $\Delta S= \Delta S_{1} +
\Delta S_{2}$.

\newsec{Setting the bosonic measure in the presence of fluxes}

In this section we will describe a connection on the gravitino and
gaugino line bundles and compute its curvature.  Without loss of
generality, we can fix attention on one boundary component, and fix
a chirality. We choose to study
 \eqn\totalline{\big( {\rm Pfaff} \hat{\Dirac}_{E_{8}}^{\partial}
\big)\otimes \big( {\rm Pfaff}
\widehat{\Dirac}_{T^{\ast}Y}^{\partial Y_{i}}
\big)^{1/2}\otimes\big( {\rm Pfaff} \widehat{\Dirac}_{\nu}^{\partial
Y_{i}} \big)^{-1/2} \otimes \big( {\rm Pfaff}
\widetilde{\Dirac}^{\partial Y_{i}} \big)^{-1/2}\to {\cal
T}^{\partial}, }
where $ {\cal T}^{\partial}$ is the space of bosonic fields on the
boundary and the   generalized Dirac operators in \totalline\ are
\eqn\gdoone{
\hat{\Dirac}_{E_{8}}^{\partial}=\Dirac_{E_{8}}^{\partial}+\gamma(\torsion),
} \eqn\gdotwo{ \widehat{\Dirac}_{T^{\ast}Y}^{\partial}=
\Dirac_{T^{\ast}Y}^{\partial}+\gamma(\torsion), } \eqn\gdothree{
\widehat{\Dirac}_{\nu}^{\partial}=
\Dirac_{\nu}^{\partial}+\gamma(\torsion), }
\eqn\gdofour{
\widetilde{\Dirac}^{\partial}= \Dirac^{\partial}-{1\over
3}\gamma(\torsion). }
where $\gamma(\cdot)$ denotes Clifford multilication by elements in
$\Omega^{\ast}(X)$, with $X:=\partial Y$.

A natural choice of connection on the determinant and Pfaffian line
bundles follows the discussion of Bismut and Freed
\bismutfreedI\bismutfreedII. Working fiberwise in ${\cal X}\to {\cal
T}^{\partial}$, we can define generalized Dirac operators
$\hat{\Dirac}$ on $X$, as the ones which appear in the definition of
the effective action, i.e., the operators \gdoone, \gdotwo,
\gdothree\ and \gdofour. We now drop the superscript $\partial$ in
the remainder of this section. The generalized Dirac operator
 \eqn\weweq{ \hat{\Dirac} = \Dirac +
\alpha_{0}\gamma(\torsion),\,\,\, \torsion\in\Omega^{3}(X), }
(where $\alpha_{0}$ is   $\alpha_{0}=1,\, -1/3$  in the case of
interest here) can   be viewed as an odd endomorphism acting on the
Hilbert bundle of spinors
\eqn\seser{ \Omega^{0}({\bf S}_{+})\oplus
\Omega^{0}({\bf S}_{-})\to {\cal T}^{\partial}, }
  where the subindices $+$ and $-$
denote the chirality of the spinor. In the Weyl basis
$\hat{\Dirac}\,$ decomposes as
 \eqn\firestt{
\hat{\Dirac}\, =\pmatrix{ 0 & \hat{\Dirac}_{-}\cr \hat{\Dirac}_{+} &
0 },}

Next, using a Riemannian structure on $\CT^{\partial}$  we can then introduce a
connection $\tilde{\nabla}$ on the Hilbert bundle $\Omega({\bf
S})\otimes\Lambda^{\ast}({\cal T}^{\partial})\to {\cal
T}^{\partial}$. This connection allows us to study the geometry of
the determinant line bundle where the effective action lives, i.e.
given the Hilbert bundle \seser\ it is possible to define its
associated determinant line bundle \eqn\detonbase{{\rm
Det}\hat{\Dirac}_+ \to {\cal T}^{\partial},} which can be also
written as\foot{See \quillen\ and \bismutfreedI, for a rigorous
definition of such infinite dimensional bundles.}

\eqn\derds{ {\rm det}\Omega^{0}({\bf S}_{+})\otimes {\rm
det}(\Omega^{0}({\bf S}_{-})^{\vee}) \to {\cal T}^{\partial}. }

This line bundle has a natural connection on it which can be
determined using heat kernel expansions \bismutfreedII. More
concretely, when restricted to a 2 dimensional submanifold
$\Sigma\hookrightarrow {\cal T}^{\partial}$, one can compute its
curvature as \bismutfreedI\bismut \eqn\curvline{\int_{\Sigma} {\cal
F}({\rm Det}\, \hat{\Dirac}_+\to {\cal T}^{\partial})=2\pi i
\int_{\pi^{-1}(\Sigma)} \Big[ {\rm Tr}_{s}a_{6}(\hat{\DiracB})
\Big]_{(12)}, } with ${\cal F}\in \Omega^2(\Sigma)$ and $\pi\colon
{\cal X}\to {\cal T}^{\partial}$ the defining fibration of the
family, with fiber $X$. \foot{The theorems of \bismutfreedII\ and
\bismut\ that we use here were stated for families of ordinary Dirac
operators and not generalized Dirac operators. However the argument
using Eq.(1.56) of \bismutfreedII, as well as the identity Eq.(5.4)
of \bismut\ can be shown to extend to the case of generalized Dirac
operators. One need only require some mild conditions on the
generalized Dirac operators, which turn out to be compatible with
the physics of our problem.} In \curvline\ we are using the heat
kernel expansion \eqn\heatexpone{ {\rm Tr}_{s}\big({\rm
exp}(-t\hat{\DiracB}^2) \big) = {{\rm Tr}_{s}a_{0}\over t^{6}}+
{{\rm Tr}_{s}a_{1}\over t^{5}}+\ldots +{\rm Tr}_{s}a_{6} + {\cal
O}(t), } where ${\rm Tr}_{s}(\,\cdot\,)={\rm
Tr}(\Gamma^{13}\cdot\,)$, and $\hat{\DiracB}$ the generalized Dirac
operator on the spin bundle of the $12$ manifold $\pi^{-1}(\Sigma)$,
defined as \eqn\gdotwl{ \hat{\DiracB} = \DiracB +
\alpha_0\Gamma(\torsion), } with $\DiracB$ the usual Dirac operator
on $\pi^{-1}(\Sigma)$, $\torsion\in \Omega^3(X)$ and
$\Gamma(\,\cdot\,)$ denotes the Clifford multiplication in the
Clifford algebra ${\rm Cliff}(12)$.

This approach allows us to compute the  curvature of the line bundle
form the  integral over two-dimensional submanifolds
$\Sigma\hookrightarrow {\cal T}^{\partial}$.

\subsec{Flux corrections to the line bundle's curvature}

If ${\rm Tr}_{s}a_{6}(\hat{\Dirac})$ is the heat kernel coefficient
associated to the generalized Dirac operator $\hat{\Dirac}$, the
curvature of the physical line bundle which appears in M-theory
\totalline, can be expressed as
\eqn\curvtotalline{\eqalign{ {\cal
F}({\cal L}_{\rm gaugino}\otimes {\cal L}_{{\rm gravitino}}\otimes {\cal
L}_{CS}\to {\cal T}^{\partial}) = {\cal F}({\cal L}_{CS}\to {\cal
T}^{\partial}) + \cr {2\pi i\over 4}\Bigg[ \int_{X} 2{\rm
Tr}_{s}a_{6}(\hat{\Dirac}_{E_{8}}) + {\rm
Tr}_{s}a_{6}(\widehat{\Dirac}_{T^{\ast}Y}^{\partial}) - {\rm
Tr}_{s}a_{6}(\widehat{\Dirac}^{\partial}_{\nu}) - {\rm
Tr}_{s}a_{6}(\widetilde{\Dirac}^{\partial})\Bigg]_{(2)}, }}
where $[\, \cdot \, ]_{(2)}$ extracts the two-form part.  Thus,
evaluating the curvature of \totalline\ is equivalent to computing
certain heat kernel coefficients.

Without evaluating the heat kernel coefficients we can make the
following observation just based on index theory. From \FM\ we know
that  the curvature \curvtotalline,   is zero for $\torsion=0$.
Since the flux can be turned on by a compact perturbation the
curvature will be an exact $2$-form on ${\cal T}^\partial$
\eqn\totalderiv{
{\cal F}({\cal L}_{\rm gaugino}\otimes {\cal L}_{{\rm gravitino}}\otimes
{\cal L}_{CS}\to {\cal T}^\partial) = {\rm d}A, }
for some {\it globally well-defined} $1$-form $A\in \Omega^{1}({\cal
T}^\partial)$.
As we have said,    ${\cal T}^\partial$ is the space of gauge
inequivalent field configurations, that is, the base of the $\CG:={\rm
Diff}(Y)\times Aut(E)$ bundle
\eqn\fibb{ 0\longrightarrow {\cal G} \longrightarrow {\rm
Met}(Y)\times {\cal A} {\buildrel \pi \over \longrightarrow} {\cal
T}^\partial \longrightarrow 0, } with ${\rm Met}(Y)$ the space of Riemannian
metrics on $Y$ and ${\cal A}$ the affine space of $E_8$-gauge
connections on the $E_8$  gauge bundle $E\to X$. We can write the
12-form $I_{12}$, used to define the curvature of the line bundle
${\cal F}=\int_X I_{12}$, as the exterior differential of a ${\cal
G}$-equivariant 11-form $I_{11}({\cal R}, F, G)$. Therefore, the
descent formalism suggests that such flux corrections do not
contribute to the anomaly.

In order to justify the above claim we proceed as follows.
As we showed above, we can construct a
generalized Dirac operator acting on the Hilbert bundle \seser. If
we now restrict to an arbitrary 2-dimensional famliy  $\Sigma\subset
\CT^\partial$ then the index of this operator, which we will denote
by ${\rm Index}\,\hat{\DiracB}$ is given by
\eqn\indexheat{{\rm Index}\,\hat{\DiracB}=\int_{\pi^{-1}(\Sigma)}
{\rm Tr}_{s}a_{6}(\hat{\DiracB}).}
One the other hand, since $\hat{\DiracB}=\DiracB + \gamma(\torsion)$
differ by a compact perturbation
\eqn\sameind{{\rm Index}\,\hat{\DiracB}={\rm Index}\,\DiracB.}
Since this applies to arbitrary families $\Sigma$ we learn that
\eqn\difinedx{\int_X {\rm Tr}_{s}a_{6}(\hat{\DiracB})=\int_X {\rm
Tr}_{s}a_{6}({\DiracB})+{\rm d}\alpha,}
for some globally well-defined $1$-form $A$ on $\CT^\partial$. However, since
the heat kernel expression is a local expression in the fields we
must have
\eqn\difinedx{ {\rm Tr}_{s}a_{6}(\hat{\DiracB})=  {\rm
Tr}_{s}a_{6}({\DiracB})+{\rm d}\alpha,}
for some $11$-form $\alpha$, that becomes zero when $\torsion=0$. In
the next section we will verify this explicitly for the case of flat
space to lowest order in $\torsion$.

\subsec{The $\IZ_2$-anomaly.}

As noted in \FM\ there is a natural real structure on the gravitino
line bundle, respected by the Bismut-Freed connection, and hence the
holonomy group is at most $\IZ_2$. In fact, it can very well be
equal to $\IZ_2$. The coupling of the gravitino to the $G$-flux
respects this real structure, and hence coupling to the $G$-flux
cannot modify the $\IZ_2$ anomaly cancellation. It will, however
change the one-loop measure. Here we give an expression for that
change.

We need to compute
\eqn\keyeq{\xi(\Dirac_{RS}+\ell^{3}\Xi\cdot G) :=
\xi\Big(\Dirac_{T^{\ast}Y}-{\ell^{3}\over 96}\flux \Big)
-\xi\Big(\Dirac+{\ell^{3}\over
288}\flux\Big)-2\xi\Big(\Dirac-{5\ell^{3}\over 96}\flux\Big).}
where $\xi$ is the invariant appearing in the APS index theorem. We
introduce a 1-parameter family of such operators by scaling $G \to t
G$ and constructing the 12-dimensional operator:
\eqn\firstope{\widehat{\DiracB}=\sigma^{2}\otimes{\partial \over
\partial t}+\sigma^{1}\otimes \Dirac +
\ell^{3}t \sigma^{1}\otimes\flux, }
acting on spinors in the twelve-manifold $Z=Y\times \IR$. In order
to apply index theory we should think of the Dirac operator as
\eqn\rewrite{\widehat{\DiracB}:=\DiracB +t \ell^{3}\Gamma(\star G),}
where $\Gamma(\star G)$ is the Clifford multiplication by $\star
G\in \Omega^{7}(Z)$ in ${\rm Cliff}(12)$, $\star$ is the
11-dimensional Hodge operator defined on $\Omega^{\ast}(Y)$, and
\eqn\diracbb{\DiracB=\sigma^{2}\otimes{\partial \over \partial
t}+\sigma^{1}\otimes \Dirac} is the Dirac operator in
$12$-dimensions. Then we have
\eqn\compui{{\partial\xi(\Dirac+\ell^{3}\flux_{t})\over \partial
t}dt = \int_{Y} {\rm Tr}_{s}\big(a_{6}(\widehat{\DiracB})
\big)_{(12)}, }
with $a_{6}$ being the $t$-independent part of the
heat kernel expansion for ${\rm exp}(-t\widehat{\DiracB}^2)$. We can
write the tensor products by the Pauli matrices in \firstope\ as
gamma matrices in $12$ dimensions. The $12$-form that we integrate
on $Y$ in \compui\ can be interpreted as the index density of
$\widehat{\DiracB}$, hence in order to extract information on the
$G$-dependence of \keyeq, we can use results from geometric index
theory for families of operators $\widehat{\DiracB}$, as we did in
the case of the local anomaly.

The index of \rewrite, is not modified by the presence of the
$G$-flux, hence  the   flux-correction to the $12$-form ${\rm
Tr}_{s}\big(a_{6}(\widehat{\DiracB}) \big)_{(12)}$ will be
\eqn\fluxcorr{ {\rm Tr}_{s}\big(a_{6}(\widehat{\DiracB})
\big)_{(12)}= {\rm Tr}_{s}\big(a_{6}(\DiracB) \big)_{(12)} +
\ell^{3}{\rm d}\varphi_{\CD} (\star G,\, {\cal R}), }
with $\int_{Y}\varphi_\CD (\star G,\, {\cal R}): {\cal T}\mapsto
\IR$ a well defined diffeomorphism-invariant function defined on the
functional space of bosonic configurations. Adding the contributions
of the various terms we obtain an expression  of the form:
\eqn\keyeqqq{
\xi\Big(\Dirac_{T^{\ast}Y}-{\ell^{3}\over 96}\flux\Big)
-\xi\Big(\Dirac+{\ell^{3}\over
288}\flux\Big)-2\xi\Big(\Dirac-{5\ell^{3}\over 96}\flux\Big)
=\xi(\Dirac_{RS}) + \ell^{3}\int_{Y}\varphi (\star G,\, {\cal R}) .}
Since $\varphi (\star G,\, {\cal R}) $ is local and gauge invariant
we see explicitly that the $\IZ_2$ anomaly cancellation is
unchanged.

\newsec{Example: Eleven manifold with flat boundaries}

Let $X:=\partial Y=\IR^{10}$ be   flat $10$-dimensional Euclidean
space.  Let $E\to X$ be the adjoint $E_{8}$-vector bundle, and
$D_{M}=\partial_{M}+{\bf A}_{M}$ the gauge connection on $E$, i.e.
$D_{M}:\, \Omega^{0}({\bf S}\otimes E)\to \Omega^{1}({\bf S}\otimes
E)$. Thus the quadratic action for the gaugino is constructed
through the generalized Dirac operator \eqn\GDO{
\hat{\Dirac}_{E_8}=\gamma^{M}D_{M}+\gamma(\torsion) } where
$\torsion=-{\ell^{3}\over 24}i_{\nu}G^{\partial}$ is the $3$-form
that comes from contracting the M-theory G-flux in the bulk $Y$,
with the normal unit vector field to the boundary $\partial Y=X$ and
\eqn\gmm{\gamma(\torsion)=\gamma^{M_{1}M_{2}M_{3}}\torsion_{M_{1}M_{2}M_{3}}.}
We consider the fibration ${\cal X}\to
{\cal T}^\partial$ encoding
the family of geometric data on the fiber $X$, i.e.  gauge
connections and fluxes, and calculate the curvature of the Pfaffian
line bundle ${\rm Pfaff}\,\hat{\Dirac}_{E_8}\to \Sigma\hookrightarrow
{\cal T}^\partial$ using \curvline\ as follows
\eqn\inieet{ {\cal F}({\rm Pfaff}\,\hat{\Dirac}_{E_8}\to \Sigma\hookrightarrow {\cal
T}^\partial) =\pi i\int_{X} {\rm Tr}_{s}a_{6}(\hat{\Dirac}_{E_8}) }
where ${\rm Tr}_{s}a_{6}(\hat{\Dirac}_{E_8})$ is the $t$-independent
finite part of the heat kernel expansion for \eqn\hke{ {\rm
Tr}_{s}{\rm exp}(-t\hat{\DiracB}_{E_8}^{2}) } when $t\to 0$ while
$t>0$. ${\rm Tr}_{s}(\cdot):={\rm Tr}(\gamma^{13}\cdot)$ means
supertrace. In contrast to the case with zero flux,
there are nonzero divergent terms in the $t\to 0$ expansion.
However, these may be easily cancelled by gauge invariant
counterterms, so we focus on the $t$-independent term.

\subsec{Determining $a_{6}$ up to ${\cal O}(\torsion^2)$}

Formally, we can expand ${\rm Tr}_{s}(a_6)$ as a series in
$\torsion$: \eqn\expandoneform{ {\rm Tr}_{s}(a_6)=\alpha_0
(\torsion) + \alpha_{1}(\torsion) + \alpha_{2}(\torsion) +
\alpha_{3}(\torsion) +\ldots, } with $\alpha_{i}(\torsion)$ a 2-form
in ${\cal T}$ which scales homogeneously under scalings of the
torsion, i.e. $\alpha_{i}(\lambda \torsion)=
\lambda^{i}\alpha_{i}(\torsion)$. For simplicity, we determine only
the lowest correction $\alpha_{1}(\torsion)$ to ${\rm Tr}_{s}(a_6)$.

In order to evaluate ${\rm Tr}_{s}a_{6}$, we are going to use known
results on heat kernel expansions for generalized Laplacians of the
type \eqn\heatres{ \Delta=-(\nabla_{N}\nabla^{N} + V) } with
$\nabla_{N}=\partial_{N}+Q_{N}$ a first order partial differential
operator, $Q_{N}{\rm d}x^{N}$ a matrix of one-forms and $V$ a scalar
matrix. For such operators, the  $t$-independent finite part of the
heat kernel expansion for \eqn\hkegeneral{ {\rm exp}\big(
-t(\nabla_{N}\nabla^{N} + V)\big) } has been calculated in flat
space using different methods, see \FHSS\ and \vandeven. Thus we
want to write $\hat{\Dirac}_{E_8}^{2}$ as an operator of the type
\heatres. If we introduce the connection
$$\nabla_{M}=\partial_{N}+
{\bf A}_{N}+3\torsion_{NM_{1}M_{2}}\gamma^{M_{1}}\gamma^{M_{2}},$$
we find
\eqn\sqgdo{\eqalign{ \hat{\Dirac}_{E_8}^{2} =
-\nabla_{N}\nabla^{N}+F_{MN}\gamma^{MN}+
\partial_{M_1}\torsion_{M_{2}M_{3}M_{4}}\gamma^{M_{1}}\gamma^{M_{2}M_{3}M_{4}} +
4\torsion^{M_{1}M_{2}M_{3}}\torsion_{M_{1}M_{2}M_{3}}. }}
 with
$F=d{\bf A}+{\bf A}\wedge {\bf A}$, the curvature of the vector
bundle $E\to X$. Hence, as $\nabla_{N}=\partial_{N}+ {\bf
A}_{N}+3\torsion_{NM_{1}M_{2}}\gamma^{M_{1}}\gamma^{M_{2}}$, $V$ in
\heatres\ is fixed to be \eqn\defx{\eqalign{
V=-F_{MN}\gamma^{MN}
-\partial_{M_1}\torsion_{M_{2}M_{3}M_{4}}\gamma^{M_{1}}\gamma^{M_{2}M_{3}M_{4}}
-4\torsion^{M_{1}M_{2}M_{3}}\torsion_{M_{1}M_{2}M_{3}}. }} Now,
having written $\hat{\Dirac}_{E_8}^{2}$ as a generalized Laplacian,
we can use the coefficient calculated in  \FHSS\vandeven,
\eqn\hcoeff{ a_{6}={1\over 6!}\Big[ V^6 + 6
V^{2}\nabla^{N}(V)V\nabla_{N}(V) + 4V^{3}\nabla^{N}(V)\nabla_{N}(V)
+ {\cal O}(V^{4}) \Big], } to evaluate the lowest order flux
correction in ${\rm Tr}_{s}a_{6}(\hat{\Dirac}_{E_8})$, neglecting
the ${\cal O}(\torsion^2)$ terms in $V$.

We now   compute   the contribution of every term in \hcoeff\ as
follows:

\noindent{$\bullet$} ${\rm Tr}_{s}\big[ V^6 \big]$. The most obvious
contribution is the leading term ${\rm Tr}(F^6)$. The first order
contribution in $\torsion$ is
\eqn\oneone{ 6{\rm Tr}_{s}\big[
\partial_{M_1}\torsion_{M_{2}M_{3}M_{4}}\gamma^{M_{1}}\gamma^{M_{2}M_{3}M_{4}}
\gamma(F)^5 \big]. }
As we are working with a $12$ dimensional Clifford algebra, only the
term proportional to $\gamma^{M_1 M_2\ldots M_{12}}$   contributes
to the supertrace in \oneone. Thus, we determine the contribution
from \oneone\ by studying the irreps of the rank 14 tensor
\eqn\tensorft{
\partial_{M_1}\torsion_{M_{2}M_{3}M_{4}}F_{M_5 M_6}\ldots F_{M_{13}M_{14}},
}
defined in dimension 12 under the group $SO(12)$. Some of the symmetries
of \tensorft\ under the permutation of indices are already known, for instance
each curvature tensor $F_{M_i M_j}$ contributes antisymmetric couples $M_i,\, M_j$,
also we know that $\torsion$ is a completely antisymmetric rank 3 tensor, etc.
A detailed analysis along these lines, shows how just the symmetric part of
$M_1$ with the triad $M_2,\, M_3$ and $M_4$ in \tensorft, gives a non zero
contribution to the supertrace \oneone. Therefore, we find
\eqn\oneoneeq{ 6{\rm Tr}_{s}\big[
\partial_{M_1}\torsion_{M_{2}M_{3}M_{4}}\gamma^{M_{1}}\gamma^{M_{2}M_{3}M_{4}}
\gamma(F)^5 \big]=6(2-12)\partial^{M}\torsion_{MM_1 M_2}
{\rm d}x^{M_1}{\rm d}x^{M_2}{\rm Tr}(F^5). }

\noindent{$\bullet$} ${\rm Tr}_{s}\big[
V^{2}\nabla^{N}(V)V\nabla_{N}(V) \big]$. Here, the $\torsion$ term
can come from the $\nabla_{N}$-derivative or from the matrix $V$.
When it comes from the
$(\nabla_{N}=D_{N}+3\torsion_{NM_{1}M_{2}}\gamma^{M_{1}}\gamma^{M_{2}})$-derivative,
with $D_N=\partial_N +{\bf A}_N$ the usual gauge differential,
we find \eqn\twoone{-6{\rm Tr}_{s}\big[ \gamma(F)^{2} \cdot
\torsion_{NM_{1}M_{2}}\gamma^{M_{1}M_2}\cdot \gamma(F)^2
D^{N}(\gamma(F)) \big],}
which is the same as
\eqn\eqppp{
-6\torsion_{M_1 M_2 M_3}{\rm d}x^{M_2}{\rm d}x^{M_3}{\rm Tr}\big(D^{M_1}(F)F^{4}\big).
}
If the $\torsion$-term comes from $V$, it cannot come from
$\partial^{N}\torsion_{NM_{1}M_{2}}\gamma^{M_{1}}\gamma^{M_{2}}$
because we would combine less than $12$ gamma matrices. Thus,
up to global numerical factors we find
\eqn\twotwo{{\rm Tr}_{s}\big[ (d\torsion)_{M_{1}M_{2}M_{3}M_{4}}
\gamma^{M_{1}M_{2}M_{3}M_{4}}\gamma(F)D^{N}(\gamma(F))\gamma(F)
D_{N}(\gamma(F)) \big]} and
\eqn\twothr{{\rm Tr}_{s}\big[
\gamma(F)^{2}D^{N}((d\torsion)_{M_{1}M_{2}M_{3}M_{4}}
\gamma^{M_{1}M_{2}M_{3}M_{4}})\gamma(F) D_{N}(\gamma(F)) \big]. }

\noindent{$\bullet$} ${\rm Tr}_{s}\big[
V^{3}\nabla^{N}(V)\nabla_{N}(V) \big]$. These terms are of the same type
as in the previous case, just differing in the order of terms. For example,
we find
 \eqn\throne{{\rm Tr}_{s}\big[
(d\torsion)_{M_{1}M_{2}M_{3}M_{4}}
\gamma^{M_{1}M_{2}M_{3}M_{4}}\gamma(F)^{2}D^{N}(\gamma(F))
D_{N}(\gamma(F)) \big], }
\eqn\vierone{
-6{\rm Tr}_{s}\big[ \gamma(F)^{3} \cdot
\torsion_{NM_{1}M_{2}}\gamma^{M_{1}M_2}\cdot \gamma(F)
D^{N}(\gamma(F)) \big],
}
etc.

\noindent{$\bullet$} ${\rm Tr}_{s}\big[ {\cal O}(V^4) \big]$. It is easy to check that
these terms only contribute to ${\cal O}(\torsion^2)$.

 \bigskip

Thus, the terms above determined are the only ones that contribute to
$\alpha_1(\torsion)$ in the expansion of the line bundle curvature in
``powers'' of $\torsion$.
Furthermore, we can group the terms in $\alpha_1(\torsion)$ that scale
as $\lambda^{5}\alpha_{1}(\torsion)$ under scalings $F\mapsto \lambda\cdot F$, of the
gauge connection curvature $F$ by $\lambda\in \IR$. This set of terms
coming from  \oneone, \eqppp\ and \vierone, can be written as
\eqn\ffive{\eqalign{
{\rm Tr}_s (V^6 + 6V^2\nabla^{N}(V)V\nabla_{N}(V) + 4V^3\nabla^{N}(V)\nabla_{N}(V) )=\cr
-60\partial^{M}\torsion_{MPQ}{\rm d}x^P {\rm d}x^Q {\rm Tr}(F^5) - 60
\torsion_{MPQ}{\rm d}x^P {\rm d}x^Q {\rm Tr}(D^{M}(F)F^4) + \ldots
}}
where the $\ldots$ refer to terms with different scaling properties.

After evaluating the other supertraces, we find the forms
$D_{N}({\rm d}\torsion)\wedge{\rm Tr}\big( D^{N}(F)F^3 \big)$ and
${\rm d}\torsion\wedge {\rm Tr}\big( F D_{N}(F) F D^{N}(F)\big)$ or
${\rm d}\torsion\wedge {\rm Tr}\big( F^2 D_{N}(F) D^{N}(F)\big)$,
depending if it comes from
${\rm Tr}_{s}\big[ V^{2}\nabla^{N}(V)V\nabla_{N}(V) \big]$
or
${\rm Tr}_{s}\big[ V^{3}\nabla^{N}(V)\nabla_{N}(V) \big]$.

Taking into account the numerical factors and recalling
 \eqn\ineet{
{\cal F}({\rm Pfaff}\,\hat{\Dirac}_{E_8}\to {\cal T}^\partial)=\pi
i\int_{X} {\rm Tr}_{s}a_{6}(\hat{\Dirac}_{E_8}), }
 we obtain
the curvature for the Pfaffian line bundle
\eqn\init{\matrix{ {\cal
F}({\rm Pfaff}\,\hat{\Dirac}_{E_8}\to {\cal T}^\partial)=-{\pi i
\over 6!}\int_{X}\Big( {\rm Tr}(F^{6})+
60\partial^{M}\torsion_{M} {\rm Tr}(F^5)
\cr
+ 60\torsion_{M} {\rm Tr}(D^{M}(F)F^4)
-20 D_{N}({\rm d}\torsion)\wedge{\rm Tr}\big( D^{N}(F)F^3 \big)\cr
-12{\rm d}\torsion\wedge {\rm Tr}\big( F D_{N}(F) F D^{N}(F)\big)
-18{\rm d}\torsion\wedge {\rm Tr}\big( F^2 D_{N}(F)
D^{N}(F)\big) \Big) + {\cal O}(\torsion^2), }}
where $F=d{\bf A} + {\bf A}\wedge {\bf A}$ and
$\torsion_{N}=\torsion_{NMP}{\rm d}x^{M}\wedge {\rm d}x^{P}$.
According to our general discussion, we expect to be able to write the
correction to the standard curvature ${\rm Tr} F^6$ as the total
derivative of a gauge invariant local expression.
In the next section we will check this explicitly.

\subsec{Writing the flux corrections as total
derivatives}

The formula \init\ allows us to compute the curvature of the
M-theory line bundle ${\cal L}_G\to {\cal T}^\partial$. The
curvature of the Chern-Simons bundle  ${\cal L}_{CS}\to {\cal
T}^\partial$ exactly cancels the $\torsion$-independent part of the
curvature. Furthermore ${\cal L}_{\rm gravitino}\to {\cal
T}^\partial$ does not contribute terms to $\alpha_0(\torsion)$ nor
$\alpha_1(\torsion)$ in flat space. Therefore only the terms in
\init\  contribute to ${\cal F}({\cal L}_G\to {\cal T}^\partial)$,
up to ${\cal O}(\torsion^2)$.

For the first correction, we work with the set of terms
\eqn\thecorr{ -60\partial^{M}\torsion_{M} {\rm Tr}(F^5)-
60\torsion_{M} {\rm Tr}(D^{M}(F)F^4), } which have  identical
behavior under scalings of $\torsion$ and $F$. An obvious candidate
to write \thecorr\ as a total divergence seems to be \eqn\candone{
-60{\rm d}\big( \torsion_{N} {\rm Tr}(F_{MP}{\rm d}x^P
F^4)\big)\delta^{NM} } with $\delta^{NM}$ the Kronecker delta, which
is the metric for the twelve dimensional space that we are dealing
with. Expanding \candone\ we find \eqn\theeeq{ {\rm d}\big(
\torsion_{N} {\rm Tr}(F_{PM}{\rm d}x^P F^4)\big)\delta^{NM}= {\rm
d}\torsion_{N} {\rm Tr}(F_{PM}{\rm d}x^P F^4)\delta^{NM} +
\torsion_{N} {\rm d}{\rm Tr}(F_{PM}{\rm d}x^P F^4)\delta^{NM}. }
Furthermore, we can write the exterior differential of a trace over
the color indices, as the trace of the covariant exterior
differential, i.e., \eqn\termqqq{ {\rm d}{\rm Tr}(F_{PM}{\rm d}x^P
F^4) = {\rm Tr}\big(D( F_{PM}{\rm d}x^P F^4) \big) } with $D={\rm
d}x^{N}D_{N}\cdot ={\rm d}x^{N}(\partial_{N}\cdot +[{\bf
A}_N,\,\cdot\,])$. The expression \termqqq\ holds because the trace
of a commutator is zero. Therefore, recalling the Bianchi identity
$DF=0$, we get \eqn\termwww{ {\rm Tr}\big(D( F_{PM}{\rm d}x^P F^4)
\big) = {\rm Tr}\big(D( F_{PM}{\rm d}x^P) F^4 \big)= {\rm
Tr}\big(D_M (F) F^4 \big), } where we have used the identity $D(
F_{PM}{\rm d}x^P)=D_M (F)$, which follows from the antisymmetry
under permutation of couples of indices in the rank 3 tensor $D_M
F_{PQ}$. Using \termwww\ we can write \candone\ as \eqn\candtwo{
{\rm d}\torsion_{N} {\rm Tr}(F_{PM}{\rm d}x^P F^4)\delta^{NM} +
\torsion_{N}{\rm Tr}\big(D_M (F) F^4 \big)\delta^{NM} } which is not
yet clearly equal to \thecorr, because the first term. To show how
\eqn\firstterm{ {\rm d}\torsion_{N} {\rm Tr}(F_{PM}{\rm d}x^P
F^4)\delta^{NM} =
\partial^{M}\torsion_{M} {\rm Tr}(F^5),
}
we study the irreps of the rank 14 tensor
\eqn\tensorsym{
\partial_{M_1}\torsion_{M_2 M_3 M_4}{\rm Tr}(F_{M_5 M_6}F_{M_7 M_8}\ldots F_{M_{13} M_{14}})
}
with the symmetries under the permutations of indices implicit in the L.H.S.
of \firstterm. These consist in the completely antisymmetry of the sets $\{M_2,\, M_3, M_4 \}$ and
$\{M_1$, $M_3$, $M_4$, $M_6$, $M_7$, $M_8$, $\ldots M_{14} \}$ and the
completely symmetry of the couple $M_2$ and $M_6$ which is to be contracted with
the symmetric tensor $\delta^{M_2 M_6}$. Taken into account these constraints, we find
that \tensorsym\ defined on a twelve dimensional space lies already in a unique irreducible
representation of $SO(12)$ on $(\IR^{12})^{\otimes 14}$. This irreducible representation
also implies the complete symmetry under permutations of the set of indices $\{M_1,\, M_2,\, M_6 \}$.
Therefore, we can write
\eqn\secoterm{\eqalign{
\partial_{M_1}\torsion_{M_2 M_3 M_4}{\rm Tr}(F_{M_5 M_6}F_{M_7 M_8}
\ldots F_{M_{13} M_{14}}){\rm d}x^{M_1 M_3 M_4 M_6 M_7\ldots M_{14}}
=\cr
\partial_{M_6}\torsion_{M_2 M_3 M_4}{\rm Tr}(F_{M_5 M_1}F_{M_7 M_8}
\ldots F_{M_{13} M_{14}}){\rm d}x^{M_3 M_4 M_1 M_6 M_7\ldots M_{14}}
}}
that after contracting with $\delta^{M_2 M_6}$ becomes identical to \firstterm\ as we wanted
to prove. Hence
\eqn\finale{
-60\partial^{M}\torsion_{M} {\rm Tr}(F^5)- 60\torsion_{M} {\rm Tr}(D^{M}(F)F^4)=
-60{\rm d}\big( \torsion_{N} {\rm Tr}(F_{PM}{\rm d}x^P F^4)\delta^{NM}\big).
}

The second correction in \init, can be written as,
\eqn\secondterm{
-20D_{N}({\rm d}\torsion)\wedge{\rm Tr}\big( D^{N}(F)F^3 \big) =
- 20\partial_{N}({\rm d}\torsion)\wedge{\rm Tr}\big(
\partial^{N}(F)F^3 \big),
}
because $D_{N}{\rm d}\torsion=\partial_{N}{\rm d}\torsion + [{\bf A}_{N},\, {\rm d}\torsion]$ and
$[{\bf A}_{N},\, {\rm d}\torsion]=0$. On the other hand, ${\rm Tr}([{\bf A}^{N},\, F]F^3)=0$, thus
\eqn\exam{\matrix{
- 20\partial_{N}({\rm d}\torsion)\wedge{\rm Tr}\big(
\partial^{N}(F)F^3 \big)=-5 \partial_{N}({\rm d}\torsion)\wedge\partial^{N}{\rm Tr}\big(
F^4 \big) = -{5\over 2}\Big[\partial_{N}\partial^{N} \Big({\rm d}\torsion\wedge{\rm Tr}\big(
F^4 \big) \Big)\cr
-\partial_{N}\partial^{N}\Big({\rm d} \torsion \Big)\wedge{\rm Tr}\big(
F^4 \big) -
{\rm d}\torsion\wedge \partial_{N}\partial^{N} \Big({\rm Tr}\big( F^4 \big) \Big)
\Big],
}}
or using the Hodge Laplacian
$\partial_{N}\partial^{N} = \star {\rm d} \star {\rm d} +
{\rm d} \star {\rm d} \star $ in Cartesian coordinates for the Euclidean space $X$, we write \exam, as
\eqn\exxam{\eqalign{
-20\partial_{N}({\rm d}\torsion)\wedge{\rm Tr}\big(
\partial^{N}(F)F^3 \big) = -{5\over 2}\Big[ {\rm d} \star {\rm d} \star
\Big({\rm d}\torsion\wedge{\rm Tr}\big( F^4 \big) \Big)\cr -{\rm d}
\star {\rm d} \star \Big({\rm d} \torsion \Big)\wedge{\rm Tr}\big(
F^4 \big) - {\rm d}\torsion\wedge {\rm d} \star {\rm d} \star
\Big({\rm Tr}\big( F^4 \big) \Big) \Big]. }} The operator $\star
{\rm d} \star {\rm d}$ never appears, because it always acts on
closed forms.

Finally, the third and fourth correction in \init, can be written
using the covariant Laplacian $D_{N}D^{N}$, as
\eqn\thirdfourth{\matrix{ -{\rm d}\torsion\wedge{\rm Tr}\big( 12 F
D_{N}(F) F D^{N}(F) + 18 F^2 D_{N}(F) D^{N}(F) \big)=\cr {\rm
d}\torsion\wedge{\rm Tr}\big(6D_{N}D^{N}(F)F^3
-3D_{N}D^{N}(F^4)+3D_{N}D^{N}(F^2)F^2 \big). }} Also, we can use a
more transparent notation, using the covariant exterior derivative
\eqn\extcov{ D={\rm d}x^{N}\wedge D_{N}={\rm d} + [{\bf
A},\,\cdot\,] } we can write the curvature $F$ as $F=D^{2}$, and the
covariant Laplacian as \eqn\covlap{ D_{N}D^{N}=\star D \star D + D
\star D \star } with $\star$ being the Hodge operator. Using the
Bianchi identity $DF=0$, we rewrite \thirdfourth, as
\eqn\tfgoon{\matrix{ {\rm d}\torsion\wedge{\rm
Tr}\big(6D_{N}D^{N}(F)F^3 -3D_{N}D^{N}(F^4)+3D_{N}D^{N}(F^2)F^2
\big)=\cr {\rm d}\torsion\wedge{\rm Tr}\big( (6D \star D \star(F)F^3
-3D \star D \star(F^4)+3D \star D \star(F^2)F^2 \big). }}
Now note that
\eqn\propone{ {\rm Tr}\big(D \star D \star
(F^{4})\big)={\rm d}{\rm Tr}\big(\star D \star (F^{4})\big) + {\rm
Tr}\big( [{\bf A},\, \star D \star (F^{4})] \big)={\rm d}{\rm
Tr}\big(\star D \star (F^{4})\big) }
is an exact form.  On the other hand, consider the $6$-forms ${\rm
Tr}\big( D \star D \star(F) F^2\big)$ and ${\rm Tr}\big(D \star D
\star(F^2) F\big)$, and differentiate them twice
\eqn\onedob{\matrix{ {\rm d}^2{\rm Tr}\big(D \star D \star(F)
F^2\big)={\rm d} {\rm Tr}\big(D[D \star D \star(F) F^2]\big)=\cr
{\rm d} {\rm Tr}\big(\star D \star(F) F^3\big)= {\rm Tr}\big(D\star
D \star(F) F^3\big) }} and \eqn\secdob{\matrix{ {\rm d}^2{\rm
Tr}\big(D \star D \star(F^2) F\big)={\rm d} {\rm Tr}\big(D[D \star D
\star(F^2) F]\big)=\cr {\rm d} {\rm Tr}\big(\star D \star(F^2)
F^2\big)= {\rm Tr}\big(D\star D \star(F^2) F^2\big) }} therefore, by
construction\ \onedob\ and \secdob\ are zero. This  means that we
can write \tfgoon, as \eqn\endthfo{\matrix{ {\rm
d}\torsion\wedge{\rm Tr}\big( 6D \star D \star(F)F^3 -3D \star D
\star(F^4)+3D \star D \star(F^2)F^2 \big) =\cr
 -3{\rm d}\torsion\wedge{\rm d}{\rm Tr}\big(
\star D \star (F^{4})
\big).
}}

Using the identities \finale, \exxam\ and \endthfo, we can write the
curvature of the M-theory line bundle as
\eqn\finalcurvature{\matrix{
{\cal F}({\cal L}_G\to {\cal T}^\partial)=
-{\pi i \over 6!}\int_{X}\Big(
60{\rm d}\big( \torsion_{N} {\rm Tr}(F_{PM}{\rm d}x^P F^4)\delta^{NM}\big) +\cr
 {5\over 2}
{\rm d} \star {\rm d} \star \Big({\rm d} \torsion \Big)\wedge{\rm
Tr}\big( F^4 \big) -{5\over 2}{\rm d} \star {\rm d} \star \Big({\rm
d}\torsion\wedge{\rm Tr}\big(F^4 \big) \Big) -{1\over 2}{\rm
d}\torsion\wedge {\rm d} \star {\rm d} \star \Big({\rm Tr}\big( F^4
\big) \Big) \Big) + {\cal O}(\torsion^2). }} This formula agrees
with the results explained in section 3, where we claimed that the
curvature of ${\cal L}_G\to {\cal T}^\partial$ is an exact form
${\rm d}A$, with $A$ being a ${\cal G}$-equivariant one form on
${\rm Met}(Y)\times {\cal A}$. From \finalcurvature, we can write
$A$ as: \eqn\oneform{\eqalign{ A = -{\pi i \over 6!}\int_{X}\Bigg(
60\big( \torsion_{N} {\rm Tr}(F_{PM}{\rm d}x^P F^4)\delta^{NM}\big) +
 {5\over 2}
 \star {\rm d} \star \Big({\rm d} \torsion \Big)\wedge{\rm Tr}\big(
F^4 \big)\cr -{5\over 2} \star {\rm d} \star \Big({\rm
d}\torsion\wedge{\rm Tr}\big(F^4 \big) \Big) -{1\over 2}{\rm
d}\torsion\wedge \star {\rm d} \star \Big({\rm Tr}\big( F^4 \big)
\Big) \Bigg) + \ldots }}
up to a globally exact form. The Hodge $\star$ depends on a metric on
$\Sigma\hookrightarrow\CT^\partial$. Note that there is a natural metric
on $\CT^\partial$, induced by the Riemannian metric itself.

\subsec{Covariant form of the Anomaly}

To get a better understanding of these flux corrections to the
anomaly, it is instructive to calculate the contribution from the
fluxes to the divergence of the gauge current using the gaussian
cutoff proposed by Fujikawa. This approach to anomaly cancellation
leads to the so-called   covariant form of the anomaly. See
\odd\bardeenzumino. Fujikawa proposed to account for the local
chiral anomaly from the variation of the measure
$[d\chi][d\bar{\chi}]$ under the action of the gauge group in the
path integral
\eqn\dfdkjfkd{  \int[d\chi][d\bar{\chi}]{\rm exp}\big(\int_X
\bar{\chi} \hat{\Dirac}\, \chi \big)  }
  If $\{ T_{a} \}$ is a basis for the Lie
algebra of the gauge group ${\cal G}={\cal E}_8$, then an
infinitesimal gauge transformation   can be expressed as $g=\II +
\Lambda^a T_a +{\cal O}(\Lambda^2)$. We can compute
\eqn\fdfdf{
{\vert d_{T_a} {\rm det}\hat{\Dirac}_{E_8}\vert\over \vert{\rm
det}\hat{\Dirac}_{E_8}\vert}:={\rm d} j_a= 2i\,{\rm
Tr}\Big[T_{a}\gamma^{11}{\rm exp}\big( -t \hat{\Dirac}_{E_8}^2 \big)
\Big], }
where $j_a \in \Omega^{9}(X)$ is the gauge current.

Of course, ${\rm Tr}\big( T_{a}\gamma^{11}{\rm exp}\big( -t
\hat{\Dirac}_{E_8}^2 \big)\big)$ must be regulated, and we do so by
taking
$$
{\rm Tr}\Big[T_{a}\gamma^{11}{\rm exp}\big( -t \hat{\Dirac}_{E_8}^2
\big) \Big],
$$
where $t=1/\Lambda$ should tend to zero. In stark contrast to the case
without fluxes, the expression for $dj_a$  has divergent terms for
$t\to 0$. These divergent terms can be shown to be total covariant
divergences of local gauge invariant expressions in the fields by a
method explained below for the $t$-independent part of the heat
kernel. Thus the current must be renormalized by adding these terms.

In order to evaluate the regulator independent part of the
supertrace \fdfdf\ we have to determine the heat kernel coefficient
$a_5$. We can use again the results of \FHSS, to calculate \fdfdf\
up to second order in $\torsion$, i.e. \eqn\gfdfgder{ {\rm d} j_a
= 2i{\rm Tr}\Big(T_a \gamma^{11}
a_5(\hat{\Dirac}_{E_8})\Big)= \beta_0(\torsion)
+\beta_1(\torsion)+\ldots,} where $\beta_k(\torsion)$ are terms that
scale homogeneously under dilations of $\torsion$, i.e. if $\lambda$
is a real parameter then $\beta_k(\lambda\torsion)=\lambda^k
\beta_k(\torsion)$.

Therefore,   the only terms in $a_5$ which   contribute up to  first
order  are
\eqn\heatjkjjk{a_5= {1\over 5!}\Big[ V^5 + 2
V\nabla_N(V)V\nabla^N(V) + 3V^2 \nabla_N(V)\nabla^N(V) + {\cal
O}(V^3)\Big]. } Doing a similar calculation as we did above for the
heat kernel coefficient $a_6$, we find \eqn\derfsdf{\eqalign{ {\rm
d} j_a = {2i \over 5!}{\rm Tr}\Big( T_a F^5 \Big)+
{4i \over 5!}\Bigg[20\partial^N\torsion_N \wedge{\rm Tr}(F^4) + 4 \torsion_N\wedge
{\rm Tr}(F D^{N}(F)F^2)
+\cr 16\torsion_N\wedge{\rm Tr}(F^3D^{N}(F))+
d\torsion\wedge {\rm Tr}\Big( T_a D_N (
F) F D^N( F) + 4 T_a F D_N( F) D^N (F) \Big) +\cr \partial_N
(d\torsion)\wedge {\rm Tr}\Big(4 T_a F^2 D^N (F) + T_a F D^N (F) F
\Big) \Bigg] + \beta_2(\torsion)+\ldots }}
where $D_N=\partial_N + {\bf A}_N$ is the gauge covariant
derivative. The ``Chern-Simons'' terms, exactly cancel the
expression ${2i \over 5!}{\rm Tr}\Big( T_a F^5 \Big)$ in
\derfsdf. We can then write the flux corrections to the anomalous
divergence of the gauge current as the covariant exterior derivative
of a gauge invariant 9-form $\Delta j (\torsion)$:
\eqn\derfsw{\eqalign{ D \Delta j = {4i \over
5!}\Bigg[20\partial^N\torsion_N \wedge F^4 + 4 \torsion_N\wedge F D^{N}(F)F^2
+16\torsion_N\wedge F^3\wedge D^{N}(F) +
d\torsion\wedge\cr \Big( D_N ( F) F D^N( F) + 4 F D_N( F) D^N
(F)\Big) + \partial_N (d\torsion)\wedge \Big(4 F^2 D^N (F) + F
D^N (F) F \Big) \Bigg] 
+\ldots }}
where we have written the expression as a Lie-algebra valued form.

We now show explicitly how this can be written as a total divergence
of a gauge invariant quantity.
For the three first terms in \derfsw, one can show how
\eqn\theq{\eqalign{
20\partial^N\torsion_N F^4 + 4 \torsion_N F D^{N}(F)F^2
+16\torsion_N F^3 D^{N}(F) =\cr
D\big( 4\torsion_N F F_{PM}{\rm d}x^{P}F^2 + 16\torsion_N F^3 F_{PM}{\rm d}x^{P}
\big)\delta^{NM}.
}}
Expanding \theq, using the identity
$D_N (F) = D (F_{MN}{\rm d}x^{M})$ gives us
\eqn\expone{\eqalign{
D\big( 4\torsion_N F F_{PM}{\rm d}x^{P}F^2 + 16\torsion_N F^3 F_{PM}{\rm d}x^{P}
\big)\delta^{NM} = \big(4{\rm d}\torsion_N F F_{PM}{\rm d}x^{P}F^2 +\cr
16{\rm d}\torsion_N F^3 F_{PM}{\rm d}x^{P})\delta^{NM}+
 4\torsion_N F D^N(F)F^2 + 16\torsion_N F^3 D^N(F),
}}
thus, we have to prove the identity
\eqn\expiden{
20\partial^N\torsion_N F^4=
\big(4{\rm d}\torsion_N F F_{PM}{\rm d}x^{P}F^2 +
16{\rm d}\torsion_N F^3 F_{PM}{\rm d}x^{P})\delta^{NM}.
}
This can be achieved by analyzing the irreps of the rank 12
tensor
\eqn\ranktwl{
\partial_{M_1}\torsion_{M_2 M_3 M_4} F_{M_5 M_6} F_{M_7 M_8}\ldots F_{M_{11}M_{12}},
}
with antisymmetry under permutations of the sets of indices $\{M_2$,
$M_3$, $M_4\}$ and $\{ M_1$, $M_3$, $M_4$, $M_5$, $M_6$, $M_7$, $\ldots M_{12}\}$
and symmetry under permutations of the couple $\{ M_2$, $M_8\}$. The only
irreducible representation of $SO(10)$ in $(\IR^{10})^{\otimes 12}$ which
satisfies such properties under permutations of indices, also verifies the
complete symmetry of the set $\{ M_1$, $M_2$ and $M_8\}$, therefore
we can prove the identity \expiden\
by using the symmetry under permutations of $M_1$ and $M_8$,
contracting $M_2$ and $M_8$ with the Kronecker delta
$\delta^{M_2 M_8}$ and contracting the other indices
with their corresponding grassmann differentials. One should also consider
the same argument with $M_{12}$ playing the role of $M_{8}$ in order
to achieve the full proof.

For the second set of terms, we realize that
$ D_N ( F) F D^N( F) + 4
F D_N( F) D^N (F)$ using the Laplacian $D_N D^N = \star D \star D +
D \star D \star$. A short calculation yields \eqn\idenlap{\eqalign{
D_N ( F) F D^N( F) + 4 F D_N( F) D^N (F) = {1\over 2}D \star D
\star(F^3) + {3\over 2}FD \star D \star(F^2) \cr - {1\over 2}D \star
D \star(F^2)F - {3\over 2}FD \star D \star(F)F - 2F^2 D \star D
\star(F). }}
If again we use the identity  $D_N (F) = D (F_{MN}{\rm d}x^{M}) :=
D(F_N)$, then we can  write \derfsw\ as
\eqn\derfesw{\eqalign{ D
\Delta j = {4i \over 5!}\Bigg[
D\big( 4\torsion_N F F_{M}F^2 + 16\torsion_N F^3 F_{M}\big)\delta^{NM} +
d\torsion\wedge \Big(  {1\over
2}D \star D \star(F^3) +\cr {3\over 2}FD \star D \star(F^2) - {1\over
2}D \star D \star(F^2)F  - {3\over 2}FD \star D \star(F)F - 2F^2
D \star D \star(F) \Big) + \cr \partial^N (d\torsion)\wedge \Big(4 F^2
D(F_N) + F D(F_N) F \Big) \Bigg] + \ldots }}
Finally, using  the Bianchi identity $DF=0$, it is easy to prove
that
\eqn\derwerfdsw{\eqalign{
\Delta j = {4i \over 5!}\Bigg[
\big( 4\torsion_N F F_{M}F^2 + 16\torsion_N F^3 F_{M}\big)\delta^{NM} +
d\torsion\wedge \Big(  {1\over
2} \star D \star(F^3) +\cr {3\over 2}F \star D \star(F^2) - {1\over
2} \star D \star(F^2)F  - {3\over 2}F \star D \star(F)F - 2F^2
 \star D \star(F) \Big) + \cr \partial^N (d\torsion)\wedge \Big(4 F^2
F_N + F F_N F \Big) \Bigg] + \ldots.
}}
This gives a non-trivial redefinition of the
gauge current, by gauge invariant flux-dependent 9-forms $\Delta j
(\torsion)$.
\bigskip

\noindent{\bf Acknowledgments}

We thank E. Diaconescu, D. Freed,  and S. Ramanujam for discussions.
This work was supported   by DOE grant DE-FG02-96ER40949.

\appendix{A}{Clifford algebras in dimensions 10, 11 and 12}

In this appendix we summarize our conventions for  the Clifford
algebras ${\rm Cliff}(n)$. (We follow  \penrspin). We follow the
Clifford algebra multiplication convention \eqn\lifford{\{
\gamma^{M}, \gamma^{N}\}=-2g^{MN}.} A natural basis for ${\rm
Cliff}(n)$, is given by the set of matrices
\eqn\ideaa{\gamma^{M_{1}M_{2}\ldots
M_{p}}=\gamma^{[M_{1}}\gamma^{M_{2}}\ldots\gamma^{M_{p}]} \quad
p=0,1,\ldots n.} For $n$ even, ${\rm Cliff}(n)$ is isomorphic to
${\rm End}({\bf S})={\bf S}^{\lor}\otimes {\bf S}$, the vector space
of endomorphisms of the spinor bundle.

For $n$ odd, there is a two to one correspondence between elements
in ${\rm Cliff}(n)$ and elements in ${\bf S}^{\lor}\otimes {\bf S}$.
This map between vector spaces is understood through the action of
the volume element $\omega$ in ${\rm Cliff}(n)$, i.e. in local
coordinates \eqn\volu{\omega = \gamma^{1}\gamma^{2}\ldots
\gamma^{n}={1\over n!}\varepsilon_{M_{1}M_{2}\ldots M_{n}}
\gamma^{M_{1}}\gamma^{M_{2}}\ldots\gamma^{M_{n}} } verifies
\eqn\volsq{\omega^{2}=(-1)^{n(n+1)/2}{\bf 1}} where ${\bf 1}$ is the
identity matrix in ${\bf S}^{\lor}\otimes {\bf S}$. As the volume
element $\omega$ commutes with every element in ${\rm Cliff}(n)$ and
the Clifford algebra is irreducible, Schur's lemma implies that
$\omega$ must be represented by   $\pm {\bf 1}$. For $n=11$, we
choose $\omega = {\bf 1}$, by convention. In local coordinates,
Clifford multiplication by the volume form $\omega$ acts as a Hodge
dual, that is,  if $H$ is a $p$-form and $\hslash$ its associated
Clifford multiplication \eqn\defHslash{\gamma (H)=\hslash =
H_{M_{1}M_{2}\ldots M_{p}}\gamma^{M_{1}M_{2}\ldots M_{p} },} then
\eqn\hodge{\omega\gamma(H)=\gamma(\star H)} with $\star$ the Hodge
star operator. Thus in odd dimensions, Clifford multiplication by a
form and by its Hodge dual are represented by the same element in
${\bf S}^{\lor}\otimes {\bf S}$.

We will also use the relation between irreducible representations of
${\rm Cliff}(2n)$ and ${\rm Cliff}(2n-1)$, i.e. if $\gamma^{M}$ is
an irrep of ${\rm Cliff}(2n-1)$, an irrep $\Gamma^{M}$ for ${\rm
Cliff}(2n)$ is given by
\eqn\firstrel{\Gamma^{M}=\sigma^{1}\otimes\gamma^{M}\qquad
M=1,\ldots, 2n-1} \eqn\secrel{\Gamma^{2n}=\sigma^{2}\otimes {\bf 1}}
\eqn\thirdrel{\Gamma^{2n+1}=\sigma^{3}\otimes {\bf 1}}
where $\sigma^{i}$ are the $2\times 2$ Pauli matrices.

\appendix{B}{Heat kernel expansions and Quantum Mechanics}

There are several algorithms to evaluate the trace ${\rm Tr}\big(
{\rm exp}(-t\hat{\DiracB}^2)  \big)$. As we show in the main text of
this paper, the coefficients associated with the expansion of such a
trace in powers of $t=1/M^2$  determine the curvature of determinant
line bundles and hence the  anomalous divergence of the gauge
current. Although there are explicit calculations of such expansions
for flat space, see \FHSS, we review here some of the techniques
used to determine such coefficients, and explain qualitatively the
one based on path integrals in supersymmetric quantum mechanics.

The main idea is to separate the interacting heat kernel
\eqn\heatkI{ \langle x \vert K(t) \vert y \rangle = \langle x \vert
{\rm exp} (-t\hat{\DiracB}^2) \vert y \rangle, } as the product of
the free heat kernel \eqn\heatkII{ \langle x \vert K_0(t) \vert y
\rangle = {1\over (4\pi t)^{n/2}} {\rm exp}\Big( -{(x-y)^2\over 4t}
\Big) } with $n$ the dimension of the $x$-space, and an interacting
part $H$ \eqn\intpa{ H(x,y;t)=\sum_{k=0}^{\infty}a_k(x,y)t^k, }
i.e., we compute \heatkI\ through the ansatz \eqn\heatkIII{ \langle
x \vert K(t) \vert y \rangle = {1\over (4\pi t)^{n/2}} {\rm
exp}\Big( -{(x-y)^2\over 4t} \Big)H(x,y;t). }
There is a large variety of algorithms to calculate the coefficients
$a_k$;  they roughly fall in three categories:

\bigskip

\noindent{$\bullet$} Recursive $x$-space algorithms based on
recursive relations among different heat kernel coefficients
\dewitt\BarVilko.

\noindent{$\bullet$} Nonrecursive algorithms based on the insertion
of a momentum basis \seeley\fujikawa.

\noindent{$\bullet$} The method of Zuk, based on
graphical representations of the heat kernel coefficients \zuk.

\bigskip

 \noindent{${\rm If}$} the supertrace of \heatkIII\ is taken,
we can evaluate the expansion using path integrals in quantum
mechanics. We can follow the ideas of \gaume, \gaumewitten\ and
\friedan, to determine the coefficients of the supertrace of the
heat kernel expansion associated to a generalized Dirac operator
$\hat{\DiracB}$ in $12$-dimensions, as the ones which appear in the
definition of the curvature of the M-theory line bundle \totalline.
Thus, given the operator $\hat{\DiracB}$, the expansion
\eqn\heatexpgdo{ {\rm Tr}_{s}\big({\rm
exp}(-t\hat{\DiracB}^2)  \big) = {{\rm Tr}_{s}a_{0}\over t^{6}}+
{{\rm Tr}_{s}a_{1}\over t^{5}}+\ldots +{\rm Tr}_{s}a_{6} + {\cal
O}(t), }
can be determined through the partition function of a supersymmetric
quantum mechanical model. The idea is to interpret \heatexpgdo\ as
the time evolution operator of a quantum mechanical system with
Hamiltonian $H=\hat{\DiracB}^2$, and calculate explicitly the
expansion \heatexpgdo, through the path integral approach to quantum
mechanics. A novelty introduced by the generalized Dirac operators
is that there are coefficients ${\rm Tr}_{s}a_{k}$ with $k<6$ which
are not zero. This differs sharply from the super heat kernel
expansions for standard Dirac operators, where the coefficients with
inverse powers of $t$ are known to be zero \foot{In \getzler\ E.
Getzler calculates index densities for generalized Dirac operators.
In his approach he introduces further scalings of the fluxes by the
regulating parameter $t=1/M^2$. The first non-vanishing term in his
alternative expansion to \heatexpgdo\ is $t$-independent. However
such scalings of the field variables are not appropriate for our
application.}.

In the standard case the vanishing of the coefficients ${\rm
Tr}_{s}a_{k}$ with $k<6$ allows us to determine ${\rm Tr}_{s}a_{6}$
by evaluating the path integral in the limit $t\to 0$. In the case
of generalized Dirac operators we find non-zero terms with inverse
powers of $t$. Thus we have to be more careful and evaluate the path
integral for a finite time interval instead of taking the limit
$t\to 0$. Path integrals in supersymmetric quantum mechanics for a
finite time interval  were analyzed in detail by
\finitetimeI\finitetimeII, and used in \finitetimeIII, to determine
index densities of generalized Dirac operators in $4$-dimensions,
which agree with the older results of \obukhov\MavromatosRU.

The type of quantum mechanical theory that we consider, is a
supersymmetric non-linear sigma model with target the
$12$-dimensional manifold $Z=\pi^{-1} ( \Sigma)$ where   $\pi :
{\cal X}\to {\cal T}^\partial$ is the projection to the space of M-theory
and $ \Sigma \hookrightarrow {\cal T}^\partial$ is  any surface  where the
curvature of ${\rm Det}\,\hat{\Dirac}\to {\cal T}^\partial$ is to be
evaluated. Here, $\hat{\Dirac}=\Dirac+\gamma(\torsion)$ stands for
any of the chiral generalized Dirac operators that couple to the
M-theory fermions at the boundary.

Therefore, let $\IR^{1\vert 1}$ denote the super Euclidean space
with one even variable and one odd variable; i.e.,
$C^{\infty}(\IR^{1\vert 1})=C^{\infty}(\IR)\otimes
\wedge^{\ast}(\IR)$. And let $\tau$ and $\theta$ be the natural even
and odd variables, respectively. We consider a quantum theory of
maps \eqn\mapa{X:\IR^{1\vert 1}\to Z} where the action we take is
\eqn\acti{{\cal S}_{SQM}= -{1\over 2}\int_{\IR^{1\vert 1}}d\tau
d\theta \Bigg\{ g_{MN}(X){dX^{M}\over d\tau }DX^{N} +
DX^{M_{1}}DX^{M_{2}}DX^{M_{3}}\torsion_{M_{1}M_{2}M_{3}}(X) \Bigg\},
} with $g_{MN}$ the metric tensor on $Z$ and $D$ is the
superdifferential \eqn\superdif{D={\partial \over
\partial\theta}-\theta{\partial \over \partial \tau}.} The
superfield that appears in \acti, can be written in a local
coordinate chart as \eqn\XXX{X^{M}=x^{M}+\theta\psi^{M}} where $x$
is a local chart for $Z$. The supersymmetry transformations are
generated by the supercharge operator $Q$ \eqn\QQQ{Q={\partial \over
\partial\theta}+\theta{\partial \over \partial \tau},} with $\delta
X^{M}=Q X^{M}$. In quantizing \acti, we construct the Hilbert space
of the theory as the space $L^{2}({\bf S}(Z))$, i.e. the space of
$L^2$-sections of the spin bundle ${\bf S}\to Z$ tensored by the
half-densities on $Z$. This space has a natural $\IZ/2\IZ$-grading
induced by the chiral decomposition ${\bf S}={\bf S}_{+}\oplus{\bf
S}_{-}\to Z$. Also, the quantum supercharge operator $Q$ is (see
\macfarlane\ for a derivation): \eqn\quantume{ Q = \hat{\DiracB} =
\DiracB + \gamma(\torsion), } which acts naturally on the quantum
Hilbert space $L^{2}({\bf S}(Z))$.

Thus, the super heat kernel expansion for $Q_{+}$ can be expressed
as the quantum mechanical partition function \eqn\parti{ {\cal Z} =
{\rm Tr}_{s}\Big[ {\rm exp}\big( -t(Q_{-}Q_{+}+Q_{+}Q_{-})/2\big)
\Big] = \int [{\rm d}X]\,{\rm exp}(-{\cal S}_{SQM}), } where we take
the action ${\cal S}_{SQM}$ defined in \acti. More concretely,
writing \acti\ in the field variables and recalling that the path
integral matches with the left-hand-side of \parti\ iff the
supercircle $X:S^{1\vert 1}\to Z$ is chosen to be a supercircle of
length $t$, we find \foot{See \finitetimeIII\ for more details.
There are terms of order ${\cal O}(t)$ which have to be included in
the action, in order to make the path integral well defined for a
finite time interval due to Weyl ordering ambiguities. Here, we just
write out the classical expression derived by expanding \acti\ in
the field variables.}
\eqn\ctionf{ {\cal S}_{SQM} = {1\over t}\int_{0}^{1}{\rm d}\tau\Big[
{1\over 2}g_{MN}{{\rm d}x^{M} \over {\rm d}\tau}{{\rm d}x^{N} \over
{\rm d}\tau} + {1\over 2}g_{MN}\psi^{M}{D \psi^{N}\over D\tau} -
{1\over 2}({\rm d}\torsion)_{MNOP} \psi^{M}\psi^{N}\psi^{O}\psi^{P}
\Big], }
where
\eqn\connfrm{ {D \psi^{N}\over D\tau} = {d\psi^N\over d \tau} +
\Gamma^{N}_{\,\,MQ} \dot x^M \psi^Q - 3 \torsion^{N}_{\,\, MQ}\dot x^M
\psi^Q}
One can use the background field approximation and expand the fields
as classical fields plus quantum fluctuations \eqn\expbf{\eqalign{
x^{M}=x_{0}^{M} + \delta x^{M} \cr \psi^{Q} = \psi_{0}^{Q} +
\delta\psi^Q. }} We are now ready to compute the path integral
\ctionf\ via a loop expansion in the parameter $t$, with $t$ playing
the role of $\hbar$. We only need compute graphs of $12^{{\rm th}}$
order in the background fermions $\psi_0$ in order to saturate the
Grassmann integration. Due to the four-fermi interaction in \ctionf,
the tree level contribution after integrating $[{\rm d}^{12} \psi_0
]$ yields terms of order ${\cal O}(t^{-3})$. For instance,
\eqn\thetermm{ -{1\over 2^3 \cdot 3! t^3} ({\rm d}\torsion)^3. }
Thus, in order to extract the full ${\cal O}(t^{0})$ contribution we
should take into account up to four-loop diagrams which are of order
${\cal O}(t^{-3})\times {\cal O}(t^3)$, since each loop order L
contributes $\CO (t^{L-1})$. In this formalism, it becomes clear how
inverse powers of $t$ appear in the expansion due to the presence of
a non-vanishing $\torsion$-flux.

 In other words, we have shown how the super heat kernel
expansion will be of the type \eqn\firstnonvanishing{ {1\over
t^3}{\rm Tr}_{s}(a_{3}) + {1\over t^2}{\rm Tr}_{s}(a_{4}) +  {1\over
t}{\rm Tr}_{s}(a_{5}) +{\rm Tr}_{s}(a_{6})+\ldots, } and we will
have to evaluate up to four-loop Feynman diagrams, in order to
determine the heat kernel coefficient ${\rm Tr}_{s}a_{6}$. Note that
five-loop Feynman diagrams are at least of order ${\cal O}(t)$ and
hence do not contribute to ${\rm Tr}_{s}a_{6}$.

Finally, we remark that if we had put in an appropriate extra
scaling in $\torsion$, as is done in \getzler\ we would have had no
divergent terms for $t\to 0$ and would have obtained the index
density
\eqn\getzlerdensity{ \int \hat A e^{{\rm d} \ltorsion} .}

\listrefs
\end